\documentclass[twocolumn,showpacs,preprintnumbers,amsmath,amssymb,superscriptaddress]{revtex4-1}% APS journal style

\usepackage{graphicx}    % Include figure files
\usepackage{latexsym}    %
\usepackage{color}       %
\usepackage{dcolumn}     % Align table columns on decimal point
\usepackage{bm}          % bold math
\usepackage{amssymb}     % for math
\usepackage{hyperref}    % add hypertext capabilities
\expandafter\let\csname equation*\endcsname\relax
\expandafter\let\csname endequation*\endcsname\relax
\usepackage{amsmath}
\usepackage{mathtools}          %loads amsmath as well

\usepackage{graphics}
\usepackage{epsfig}
\usepackage{subfigure}
\usepackage{nicefrac}
\usepackage{verbatim}
\usepackage{enumerate}
\usepackage{footnote}
\usepackage{booktabs}
\usepackage{siunitx}

\begin{document}

\title{Multiscale velocity correlations in turbulence and Burgers turbulence: Fusion rules, Markov processes in scale, and multifractal predictions}

\author{Jan Friedrich}
\affiliation{Institut f\"ur Theoretische Physik I, Ruhr-Universit\"at Bochum, Universit\"{a}tsstra\ss{}e 150, 44780 Bochum, Germany}
\author{Georgios Margazoglou}
\affiliation{Dept. Physics and INFN, University of Rome ``Tor Vergata'',
Via della Ricerca Scientifica 1, I-00133 Roma, Italy}
\affiliation{Computation-based Science and Technology Research Center, Cyprus Institute, 20 Kavafi Street, 2121 Nicosia, Cyprus}
\author{ Luca Biferale}
\affiliation{Department Physics and INFN, University of Rome ``Tor Vergata'',
Via della Ricerca Scientifica 1, 00133 Roma, Italy}
\author{Rainer Grauer}
\affiliation{Institut f\"ur Theoretische Physik I, Ruhr-Universit\"at Bochum, Universit\"{a}tsstra\ss{}e 150, 44780 Bochum, Germany}

\begin{abstract}
We compare different approaches towards an effective description of multiscale velocity field correlations in turbulence. Predictions made by the operator product expansion, the so-called fusion rules, are placed in juxtaposition to an approach that interprets the turbulent energy cascade in terms of a Markov process of velocity increments in scale. We explicitly show that the fusion rules are a direct consequence of the Markov property provided that the structure functions exhibit scaling in the inertial range. Furthermore, the limit case of joint velocity gradient and velocity increment statistics is discussed and put into the context of the notion of dissipative anomaly.  We generalize a prediction made by the multifractal model derived by Benzi \textit{et al}. [Benzi \textit{et al}.,Phys. Rev. Lett. \textbf{80}, 3244 (1998)] to correlations among inertial range velocity increment and velocity gradients of any order.  We show that for the case of squared velocity gradients such a relation can be derived from first principles. Our results are benchmarked by intensive direct numerical simulations of Burgers turbulence.\footnote{\textbf{This is a postprint version of the article published in Phys. Rev. E 98, 023104 (2018)}}
\end{abstract}


\maketitle
\section{Introduction}
\label{sec:intro}
Three-dimensional turbulence is a paradigmatic out-of-equilibrium system with connections to fundamental questions in statistical mechanics~\cite{frisch:1995,monin} and many other applied problems in different disciplines, e.g., mechanical engineering~\cite{Davidson2015}, atmospheric physics~\cite{Wyngaard2010}, geophysics~\cite{Pope2001},
and astrophysics~\cite{Parker1979}. One of the most striking features of turbulence is that, already when stirred with a Gaussian, homogeneous, and isotropic forcing, the flow develops highly nontrivial, non-Gaussian, and multiscale statistical properties in the limit of high Reynolds numbers. Here the Reynolds number is the control parameter that defines the relative intensity of nonlinear vs linear terms in the Navier-Stokes equation
\begin{equation}
  \frac{\partial}{\partial t} {\bf v}({\bf x},t)
  + {\bf v}({\bf x},t) \cdot \nabla {\bf v}({\bf x},t) =
  -\nabla p({\bf x},t)+ \nu \nabla^2 {\bf v}({\bf x},t)\;.
  \label{eq:ns}
\end{equation}
The existence of anomalous scaling properties goes under the name of intermittency, which is empirically found in all three-dimensional turbulent flows in nature and is still lacking a clear understanding and derivation from the underlying equations of fluid motion. Accordingly, this phenomenon of small-scale intermittency manifesting itself, e.g., in the form of the non-self-similarity of the probability density function (PDF)  of
longitudinal velocity increments
\begin{equation}
 \delta_r v= \left[{\bf v}({\bf x}+{\bf r})-{\bf v}({\bf x})\right]\cdot \frac{{\bf r}}{r} \quad \textrm{for} \quad r>0\;,
\label{eq:inc}
\end{equation}
is still one of the most compelling experimental, numerical and theoretical open problems of fully developed turbulence.
Many studies of turbulence research have been devoted to the experimental and theoretical examination of the scaling exponents $\zeta(n)$ of structure functions $\langle (\delta_r v)^{n}\rangle \sim r^{\zeta(n)}$ in the inertial range~\cite{frisch:1995}. Here, Kolmogorov's phenomenological description of the turbulent energy cascade, i.e., the transport process of energy from large to small scales, predicts $\zeta(n)=n/3$, which in turn implies a self-similar velocity increment PDF. The effects of intermittency lead to deviations from Kolmogorov's theory and $\zeta(n)$ has been empirically found  to be a nonlinear function of $n$~
\cite{frisch:1995,Yakhot2006,Ishihara2009,Benzi2010,Sinhuber2017,Iyer2017,Reinke2017}.

The pivotal role of the turbulent energy cascade in turbulence theory immediately suggests the importance of extending the  analysis based on single-scale observables (\ref{eq:inc}) to multiscale velocity increments, which should also lead to a better understanding of local and nonlocal correlations inside the inertial range and among inertial and viscous scales. Owing to the prohibitive analytical difficulties to attack the Navier-Stokes equation (\ref{eq:ns}), the attention has been also often focused on other dynamical models of turbulence, in particular to the Burgers equations, a simplified one-dimensional and compressible version of the Navier-Stokes equation. Here the only nonlinearity enters through the advective term
\begin{equation}
  \frac{\partial }{\partial t} v(x,t) + v(x,t) \frac{\partial}{\partial x} v(x,t)
  = \nu \frac{\partial^2}{\partial x^2}v(x,t)\;.
\label{eq:burgers_intro}
\end{equation}
It is well known that the Burgers equation develops a quasishock for generic smooth initial conditions, a property that is also connected to
anomalous scaling of the velocity increments~\cite{Bec2007}.
Furthermore, in this paper we impose periodic boundary conditions and deal only with the forced Burgers case  (see Sec.~\ref{sec:burgers}) which can also be treated by using the Hopf-Cole transformation~\cite{Bec2007}. Neglecting the forcing contributions would make the problem exactly solvable, however, the introduction of suitable boundary conditions can change the problem considerably.

In the following we will address both the Navier-Stokes and Burgers equation using different statistical approaches to describe their multiscale correlation properties, together with a series of quantitative validations using direct numerical simulations of Eq.~(\ref{eq:burgers_intro}). In particular, we will
compare the two seemingly different approaches of the operator-product expansion~\cite{Eyink1993,Lvov1996,Lvov1996a,Benzi1998,Benzi1999} and the Kramers-Moyal approach
~\cite{Friedrich1997,Friedrich2018,Friedrich2016b,Friedrich2017}. It will be shown that both methods yield the same predictions for multiscale velocity increment correlations, the so-called fusion rules. Subsequently, we will address the case where one of the increments matches the velocity gradient within the framework of the multifractal approach
~\cite{frisch:1995,Benzi1998,Frisch1991}. We will prove a particular expression of the multifractal (MF) approach from first principles in Burgers turbulence, i.e., by deriving an exact velocity increment hierarchy from the Burgers equation.

Historically, one of the first multiscale analyses in turbulence was carried out in~\cite{Eyink1993} where the operator-product expansion from quantum field theory~\cite{Wilson1969} was invoked. In this framework, one can
derive the relation for the two-increment (three-point) quantity
\begin{equation}
  \langle (\delta_r v)^p (\delta_{R} v)^q \rangle
  \sim \frac{\langle (\delta_r v)^p\rangle}{\langle (\delta_{R} v)^p\rangle} \langle (\delta_{R} v)^{p+q}\rangle\;
  \label{eq:fusion}
\end{equation}
for $\eta < r \le R  \le L$, where $\eta$ is the dissipation scale and $L$ the integral length scale. Moreover, we assume that one of the two extremes of the interval of length $r$ and $R$ coincides and that both increments are collinear.
These relations are known as fusion rules and they have been  analyzed both theoretically and numerically~\cite{Lvov1996,Lvov1996a,Benzi1998,Benzi1999}. It should be noted that the fusion rules necessarily imply a reduction of the spatial complexity of the problem: The three-point quantity on the left-hand side of Eq.~(\ref{eq:fusion}) can be cast in terms of two-point quantities, the structure functions
$\langle (\delta_{r} v)^n\rangle$. For three-dimensional isotropic and homogeneous turbulent flows, one can show~\cite{Hill2001} that the most general tensorial two-point velocity correlation function can always be decomposed in terms of longitudinal or transverse velocity structure functions. Here, for the sake of simplicity, we will always limit the discussion to the case when all distances are collinear with the velocity increments taken on the longitudinal direction as given by Eq.~(\ref{eq:inc}). Furthermore, this is the only possible case for  one-dimensional Burgers turbulence (discussed below).

In the following, we will address the multiscale correlation function (\ref{eq:fusion}) by using the MF  model~\cite{frisch:1995,Benzi1998,Frisch1991} as well as the Kramers-Moyal (KM) approach
~\cite{Friedrich1997,Friedrich2018,Friedrich2016b,Friedrich2017} in order to describe the evolution of velocity increment PDFs across the inertial range. Within the MF model we will also address multiscale correlation functions when one of the velocity increment is calculated at fused points, i.e., when the increment is smaller than the viscous dissipative cutoff. The latter case is important to discuss in the context of the so-called
dissipative anomaly~\cite{Polyakov1995a} that emerges in a multi-point PDF hierarchy of Burgers turbulence (see also the discussion in Sec.~\ref{sec:burgers} of this paper). Let us mention that there exist different definitions for dissipative anomaly in the literature, both connected to local or averaged quantities \cite{frisch:1995, Polyakov1995a,Duchon2000}; in this paper we are only interested in the definition in terms of averaged quantities and in the limit of small but nonzero viscosity. We note that these different definitions address the same physical issue as already noted by Polyakov [see equation and discussion after Eq.~\eqref{eq:zeta_q} in \cite{Polyakov1995a}]. Most of the theoretical arguments are general and can be applied both to the three-dimensional homogeneous and isotropic Navier-Stokes equation and to the one-dimensional Burgers equation. We will then present a series of detailed numerical benchmarks for the latter case only, where one can achieve a separation of scales large enough to make precise quantitative statements. The paper is organized as follows. In Sec.~\ref{sec:fusion} we outline the usual derivation of the fusion rules (\ref{eq:fusion}) and discuss the dissipative cutoff within the framework of the MF model. Henceforth, it will be shown  in Sec.~\ref{sec:markov_fusion} that the fusion rules (\ref{eq:fusion}) can be derived from
the KM expansion associated with a Markov process~\cite{Risken}.
Sec.~\ref{sec:burgers} contains a derivation of a multi-increment PDF hierarchy from the Burgers equation which leads to a validation of the MF prediction from first principles. In Sec.~\ref{sec:dns} we will examine both fusion rules and the MF predictions in direct numerical simulations of Burgers turbulence.

\section{Fusion-rules and the multifractal model}
\label{sec:fusion}
The derivation of  the fusion rules (\ref{eq:fusion}) starts from the assumption that the small-scale statistics
of $\delta_r v$ is related to the large-scale configuration $\delta_{R} v$
via the multiplier $\lambda(r,R)$ according to
\begin{equation}
 \delta_r v \sim \lambda(r,R) \delta_{R} v.
 \label{eq:multiplier}
\end{equation}
Furthermore, we assume that $\lambda(r,R)=\lambda(r/R)$, which is a consequence of a purely uncorrelated multiplicative process in addition to homogeneity along the energy cascade~\cite{Benzi1998,Benzi1999} and yields
\begin{eqnarray}\nonumber
    \langle (\delta_r v)^p (\delta_{R} v)^q \rangle
    &\sim& \langle \lambda(r/R)^p  (\delta_{R} v)^{p+q} \rangle\\
    &\sim& \langle \lambda(r/R)^p \left[\lambda(R/L) (\delta_{L} v)\right]^{p+q} \rangle \;,
\end{eqnarray}
where we required that the large-scale increment is related to the integral scale increment by the same relation (\ref{eq:multiplier}). Furthermore, $
\delta_{L} v$ is assumed to be statistically independent of the multiplier
$\lambda(r/L)$, which yields
\begin{equation}
  \langle (\delta_r v)^p (\delta_{R} v)^q \rangle
  \sim \langle \lambda(r/R)^p \lambda(R/L)^{p+q} \rangle \langle(\delta_{L} v)^{p+q} \rangle\;,
  \label{eq:intermediate_fusion}
\end{equation}
but also implies that $\langle (\delta_r v)^p \rangle =
\langle(\delta_{L} v)^p \rangle \langle \lambda(r/L)^p \rangle$. Hence, in the high-Reynolds number limit ($\textrm{Re}=\sqrt{\langle {\bf v}^2 \rangle}L / \nu  \gg 1$, with the kinematic viscosity $\nu$) where we expect scaling of the structure functions
$\langle (\delta_r v)^p \rangle \sim (r/L)^{\zeta(p)}$, we can demand that
$\langle \lambda(r/R)^p \rangle \sim (r/R)^{\zeta(p)}$.
The last hypothesis that enters the derivation of the fusion rules (\ref{eq:fusion}) is that the multipliers obey an uncorrelated multiplicative process, which allows the splitting of the first expectation value
on the right-hand side of Eq.~(\ref{eq:intermediate_fusion})
\begin{eqnarray}\nonumber
   \langle (\delta_r v)^p (\delta_{R} v)^q \rangle
    &\sim& \underbrace{\langle \lambda(r/R)^p \rangle}_{(r/R)^{\zeta(p)}} \underbrace{\langle \lambda(R/L)^{p+q} \rangle
    \langle(\delta_{L} v)^{p+q} \rangle}_{\langle \lambda(R/L)^{p+q}(\delta_{L} v)^{p+q} \rangle}\\
    &\sim& \frac{\langle(\delta_r v)^p\rangle}{\langle (\delta_{R} v)^p\rangle} \langle (\delta_{R} v)^{p+q}\rangle\;.
\end{eqnarray}
In the following, we will also consider the case when the small-scale increment in Eq.~(\ref{eq:fusion}) approaches the velocity gradient. On the basis of the MF model one can deduce the existence of an intermediate dissipation range~\cite{Paladin1987}, corresponding to a continuous range of dissipation lengths $\eta(h,\nu)$, where $h$ denotes the continuous range of scaling exponents of the MF model
(see also~\cite{Frisch1991} and note that the MF model is also contained in Mellin's transform in combination with the method of steepest descent~\cite{Yakhot2006}). In addition, the MF model can be invoked in order to investigate the Reynolds number dependence of moments of velocity derivatives~\cite{Nelkin1990}. By the use of these multifractal calculations in combination with the
intermediate dissipation range cutoff, one can derive
expressions for joint velocity gradient-increment statistics~\cite{Benzi1998,Benzi1999} such as
\begin{equation}
  \left \langle \left(\frac{\partial v(x)}{\partial x}\right)^2
  [\delta_{R} v(x)]^q \right \rangle \sim \frac{R^{\zeta(q+3)-1}}{\nu}.
  \label{eq:diss_fusion}
\end{equation}
Here we explicitly wrote the dependence of the increment $\delta_{R}v$ on $x$ in order to indicate that the velocity gradient and the velocity increment are calculated with one point in common, $x$. Moreover, it must be stressed that this relation only holds if the scaling exponents fulfill Kolmogorov's 4/5 law, i.e., $\zeta(3)=1$.

We now want to generalize the previous expression (\ref{eq:diss_fusion})
to arbitrary orders of the velocity gradient.
To this end, we define the quantity
$D_{p,q}(\nu,R)=\langle (\partial v/\partial x)^p (\delta_R v)^q \rangle$,
which can be written in terms of the dissipative scale $\eta(\nu)$ as
\begin{equation}
 D_{p,q}(\nu,R)= \left \langle \left(\frac{\delta_{\eta} v}{ \eta}\right)^p
  (\delta_{R} v )^q \right \rangle.
  \label{eq:diss_fusion_generalized_a}
\end{equation}
The MF ansatz is based on the introduction of a set of scaling exponents $h$, so there exists a local scaling law
\begin{equation}
\delta_{\eta}v = (\eta /R)^h \delta_R v,
\label{eq:mf1}
\end{equation}
with probability $P_h(\eta,R)=(\eta/R)^{3-D(h)}$, where $D(h)$ is the fractal dimension of the set and where the velocity increment is H\"older continuous with exponent $h$ (see also \cite{frisch:1995}). Furthermore, the dissipative scaling is defined by requiring an $O(1)$ local Reynolds number \cite{Paladin1987}
\begin{equation}
\textrm{Re}_{\textrm{loc}}=\frac{\eta \delta_{\eta} v}{\nu}  \sim O(1)\;.
\label{eq:mf2}
\end{equation}
As a result,  we get a fluctuating $\eta$  which depends on $h$ and $\nu$.
Using (\ref{eq:mf1}) and (\ref{eq:mf2}) in (\ref{eq:diss_fusion_generalized_a}), we obtain the first conditional expectation
\begin{eqnarray}\nonumber
  &\left \langle \left(\frac{\delta_{\eta} v}{ \eta}\right)^p (\delta_{R} v )^q \Big| \delta_R v \right \rangle \\ \nonumber
  &\sim \int \textrm{d}h (\delta_R v)^{q+p} \, R^{-p} \, \left(\frac{\nu}{R\delta_R v} \right)^{[p(h-1) +3-D(h)]/(1+h)}  \\
 &\sim \frac{(\delta_R v )^{[q+p+\phi(p)]}R^{\phi(p)-p}}{\nu^{\phi(p)}}\;,
 \label{eq:diss_fusion_generalized}
\end{eqnarray}
where we have used a saddle-point estimate in the limit of infinite Reynolds numbers $\nu \to 0$ in order to get the exponent
\begin{equation}
\phi(p) = - \min_h{\frac{p(h-1) +3-D(h)}{1+h}}.
\end{equation}
Finally, we can estimate the unconditioned expectation value by considering again the MF ansatz to connect the velocity increment at scale $R$ with the large-scale velocity fluctuation $v_L$,
\begin{equation}
\label{eq:multi}
\delta_R v = (R/L)^h \delta_L v,
\end{equation}
and integrating over all possible $h$,
\begin{equation}
\label{eq:multi2}
D_{p,q}(\nu,R) \sim \int \textrm{d}h  R^{3-D(h)} \frac{(\delta_R v )^{[q+p+\phi(p)]} R^{\phi(p)-p}}{\nu^{\phi(p)}},
\end{equation}
where we have taken $L=1$ for simplicity.  Plugging (\ref{eq:multi}) in (\ref{eq:multi2}) and using
again a saddle-point estimate in the limit $ R \ll L=1$, we get
 \begin{equation}
  D_{p,q}(\nu,R) \sim \textrm{Re}^{\phi(p)} \, R^{\zeta(p+q+\phi(p))}R^{\phi(p)-p}\;,
  \label{eq:diss_anomaly_mf}
\end{equation}
where the viscosity from Eq.~(\ref{eq:multi2}) has been replaced by the dimensionless Reynolds number $\textrm{Re}$ for which the relation $\textrm{Re} \sim O(1) / \nu$ holds. The exponents $\zeta(q)$ are the scaling exponents of the structure function of order $q$,
  \begin{equation}
  \langle (\delta_R v )^q \rangle \sim \int \textrm{d}h (\delta_R v)^q R^{3-D(h)}   \sim R^{\zeta(q)}\;,
\end{equation}
  with
\begin{equation}
  \zeta(q) = \min_h{[qh+3-D(h)]}\;.
  \label{eq:zeta_q}
\end{equation}
It is important to remark that within the MF ansatz the scaling exponents of the velocity gradient, i.e., $\left \langle \left(\partial v/\partial x\right)^p \right \rangle \sim \textrm{Re}^{\phi(p)}$, and the structure function scaling exponent are connected via~\cite{frisch:1995,Nelkin1990}
\begin{equation}
  \phi(p)=[q-\zeta(q)]/2 \qquad \textrm{and} \qquad p=[\zeta(q)+q]/2\;.
  \label{eq:multi_relation}
\end{equation}
Using this expression, it is easy to see that, provided the third-order single-scale structure function satisfies
the 4/5 law $\zeta(3)=1$, then
for $p=2$ the expression (\ref{eq:diss_anomaly_mf})
possesses the remarkable property that it is inversely  dependent on the viscosity $\nu$, e.g., $\nu \left \langle \left[\partial v(x)/\partial x\right]^2 [\delta_{R} v (x)]^q \right \rangle$ remains a finite quantity in the limit $\nu \rightarrow 0$, which is a sort of generalized  dissipative anomaly \cite{frisch:1995}.

In Sec.~\ref{sec:burgers} we will prove Eq.~(\ref{eq:diss_fusion}) from
first principles in Burgers turbulence and discuss the effects of pressure contribution that we have to face in the more general case of three-dimensional Navier-Stokes equation. A different approach to the turbulent velocity gradient statistics  was carried out recently~\cite{yakhot:2001,Yakhot2004,Yakhot2007}. Here a series of order-dependent  dissipative scales $\eta_{2n}$  is introduced starting from  a balancing of inertial and diffusive terms of the equation for the $2n$th-order longitudinal structure function
\begin{equation}
  \eta_{2n}= L \textrm{Re}^{1/[\zeta(2n)-\zeta(2n+1)-1]}.
  \label{eq:victor2}
\end{equation}
Furthermore, the moments of the velocity gradient can be related to the
structure functions via the local dissipation Reynolds number (\ref{eq:mf2}) according to
\begin{equation}
  \left \langle \left| \frac{\partial v}{\partial x} \right|^n \right \rangle
  \approx \left \langle \left| \frac{\delta_{\eta}v}{\eta} \right|^n \right \rangle = \frac{\left \langle(\delta_{\eta}v)^{2n} \right \rangle}{\nu} \sim \textrm{Re}^n \eta_{2n}^{\zeta(2n)} ,
\end{equation}
Equation~(\ref{eq:victor2}) implies Reynolds number scaling of the velocity gradients according to
\begin{equation}
    \left \langle \left| \frac{\partial v}{\partial x} \right|^n \right \rangle
    = \textrm{Re}^{s_n}\;,
    \label{eq:yakhot_gradient}
\end{equation}
where
\begin{equation}
  s_n = n+ \frac{\zeta(2n)}{\zeta(2n)-\zeta(2n+1)-1}\;.
  \label{eq:yakhot_prediction}
\end{equation}
Here $\zeta(n)$ denotes the exponent of absolute values of structure functions $\left \langle |\delta_r v|^n \right \rangle$.
The above prediction is different from the MF result for $q=0$ in Eq.~(\ref{eq:diss_anomaly_mf}) (see also~\cite{Benzi2009} for a quantitative comparison). Furthermore, it is not obvious how Eq.~(\ref{eq:yakhot_gradient}) should be  generalized in order to predict the multiscale dissipative-inertial correlation function (\ref{eq:diss_anomaly_mf}).
It is important to remark that the above relation pertains only to absolute value velocity increments.  If one extends it to the signed quantities,
$\left (\langle \delta_r v)^n \right \rangle$, it would be  inconsistent with the existence of a dissipative anomaly, i.e. with the constraint $s_2=1$, unless the relation $\zeta(5)-\zeta(4) = \zeta(4)-1$ holds.
Inserting $\zeta(3)=1$, this relation suggests mono-scaling $\zeta(5)-\zeta(4) = \zeta(4)-\zeta(3)$, which is at odds with intermittency effects observed in three-dimensional turbulence (but compatible with the Burgers scaling, discussed below).

Nevertheless, the MF model must yet be considered as the only description of multiscale correlations in turbulence capable of reproducing the existence of dissipative anomaly. The latter depends only on the requirement that the exact 4/5-law is satisfied in the inertial range, i.e., $\zeta(3)=1$.

\section{Markov property in scale and fusion rules}
\label{sec:markov_fusion}
Another description of multi-increment statistics in turbulence was proposed in~\cite{Friedrich1997}, using a Markov process of velocity increments in scale for the turbulent energy cascade. It is worth specifying further the concept of this cascade process. In three-dimensional turbulence, the vortex stretching term induces small-scale structures which are believed to be vortex tubes or vortex sheets~\cite{Yeung2015}. In its original form~\cite{frisch:1995}, the turbulent energy cascade suggests that this destabilization of large-scale vortical structures is accompanied by an energy transfer from large to small scales. This particular interpretation of the turbulent energy cascade concentrates on the geometrical structures inherent in the particular flow and may differ in other types of flows, e.g., Rayleigh-B\'enard convection (plumes), magnetohydrodynamic turbulence (current sheets), pipe flows (boundary layer), and finally shocks in Burgers turbulence. In the following,
we are concerned with a stochastic description of the energy transport across scales without paying attention to the underlying structures. The latter approach starts from the definition of the $n$-increment PDF
\begin{equation}
  f_n(v_n,r_n;v_{n-1},r_{n-1};\ldots;v_1,r_1)= \prod_{i=1}^n
  \left \langle \delta(v_i - \delta_{r_i}v) \right \rangle,
  \label{eq:n_inc_pdf}
\end{equation}
where we restricted ourselves to longitudinal velocity increments (\ref{eq:inc}) only (note that the inclusion of mixed longitudinal and transverse increment statistics necessarily complicates the entire procedure~\cite{Siefert2004}).
According to Bayes' theorem, we can define the conditional probabilities
\begin{equation}
  p(v_3,r_3|v_2,r_2;v_1,r_1)=\frac{f_3(v_3;r_3,v_2,r_2;v_1,r_1)}{f_2(v_2,r_2;v_1,r_1)}
  \label{eq:bayes1}
\end{equation}
and
\begin{equation}
  p(v_2,r_2|v_1,r_1)=\frac{f_2(v_2,r_2;v_1,r_1)}{f_1(v_1,r_1)}.
  \label{eq:bayes2}
\end{equation}
Henceforth, the localness of interactions of the cascade process of the longitudinal velocity increments in scale is ensured by the Markov property in scale
\begin{equation}
  p(v_3,r_3|v_2,r_2;v_1,r_1)=p(v_3,r_3|v_2,r_2),
  \label{eq:markov}
\end{equation}
where we assume that $\eta < r_3 \le r_2 \le r_1 < L$.
The Markov property implies a considerable reduction of the spatial complexity of the velocity increment statistics, which can be deduced from the $n$-increment PDF (\ref{eq:n_inc_pdf}): If one imposes the scale ordering $\eta < r_n \le r_{n-1} \le \ldots \le r_1 < L$, this ($n+1$)-point quantity factorizes due to the Markov property according to
\begin{eqnarray}
  &~& f_n(v_n,r_n;v_{n-1},r_{n-1};\ldots;v_1,r_1)\\ \nonumber
  &=& p(v_n,r_n|v_{n-1},r_{n-1}) \times\cdots \times p(v_2,r_2|v_1,r_1) f_1(v_1,r_1)\;,
\end{eqnarray}
Hence, the Markov property constitutes a three-point-closure of the multi-increment statistics~\cite{Stresing2010a,Friedrich2017}.

In the following, we examine the implications of  (\ref{eq:markov}) for the multiscale moments (\ref{eq:fusion}). A central notion of a Markov process is that the transition PDF follows the same KM expansion as the one-increment PDF~\cite{Risken}, namely
\begin{eqnarray}
 -\frac{\partial}{\partial r_2}  f_1(v_2,r_2) &=&\hat L_{\rm KM}(v_2,r_2) f_1(v_2,r_2)\;,
  \label{eq:km_exp}\\
 -\frac{\partial}{\partial r_2}  p(v_2,r_2|v_1,r_1) &=&\hat L_{\rm KM}(v_2,r_2) p(v_2,r_2|v_1,r_1)\;, \qquad
    \label{eq:km_exp_trans}
\end{eqnarray}
where the KM operator is defined as
\begin{equation}
  \hat L_{\rm KM}(v_2,r_2)=\sum_{k=1}^\infty (-1)^k \frac{\partial^k}{\partial v_2^k}  D^{(k)}(v_2,r_2)\;.
\end{equation}
Furthermore, the minus sign in Eq.~(\ref{eq:km_exp_trans}) indicates that the process occurs from large to small scales and the KM coefficients are defined as
\begin{equation}
 D^{(k)}(v_2,r_2)=\frac{1}{k!} \lim_{r_3 \rightarrow r_2}
 \int \textrm{d}v_3 \frac{(v_3-v_2)^k}{r_2 -r_3} p(v_3,r_3|v_2,r_2)\;.
 \label{eq:KMcoeff}
\end{equation}
The KM expansion (\ref{eq:km_exp}) allows for an appealing formulation of intermittency via an evolution of the one-increment PDF (\ref{eq:km_exp}) in scale.
Moreover, scaling solutions for the structure functions, i.e.,
$\langle (\delta_r v)^n \rangle \sim r^{\zeta(n)}$ necessarily imply KM coefficients of the form~\cite{Friedrich2016b,Friedrich2018,Nickelsen2017}
\begin{equation}
  D^{(k)}(v_2,r_2)= \frac{(-1)^k K_k}{k!} \frac{v_2^k}{r_2},
  \label{eq:KM_scaling}
\end{equation}
as can be seen by taking the moments $\int \textrm{d}v_2\; v_2^n f(v_2,r)=\langle (\delta_r v)^n \rangle$ from Eq.~(\ref{eq:km_exp})
and setting $r_2=r$,
\begin{equation}
  -\frac{\partial}{\partial r}\langle (\delta_r v)^n \rangle=
  \sum_{k=1}^n {n \choose k} K_k(-1)^k \frac{\langle \delta_r v^n \rangle}{r},
\end{equation}
Dividing by the structure function of order $n$ yields
\begin{equation}
  -\frac{\partial}{\partial r}\ln \langle (\delta_r v)^n \rangle=\frac{1}{r}
  \sum_{k=1}^n {n \choose k} K_k(-1)^k\;.
\end{equation}
Integrating this equation from $r$ to $L$ yields
\begin{equation}
  \langle (\delta_r v)^n \rangle=\langle (\delta_L v)^n \rangle \left(\frac{r}{L}\right)^{-\sum_{k=1}^n {n \choose k} K_k(-1)^k}\;.
  \label{eq:km_mom}
\end{equation}
Accordingly, the reduced KM coefficients $K_k$ are related to the scaling exponents $\zeta(n)$ according to
\begin{equation}
 \zeta(n) = - \sum_{k=1}^n {n \choose k} K_k (-1)^k\;.
 \label{eq:zeta}
\end{equation}
All currently known phenomenological models of turbulence are reproduced by a suitable choice of the reduced KM coefficients listed in Table~\ref{table:KM}. Another important implication of this KM description of structure function scaling follows directly from the moment solution (\ref{eq:km_mom}): In order to obtain nonvanishing odd order moments (such as Kolmogorov's 4/5 law $\langle \delta_r v^3 \rangle=-\frac{4}{5}\langle \varepsilon \rangle r$ ) at a scale $r$ one must have non-vanishing odd order moments at large scales $L$. In other words, the symmetric form of the KM expansion dictated by the coefficients (\ref{eq:KM_scaling}) is not able to generate skewness during the cascade process; it can only transport an initial large-scale skewness in the PDF down in the cascade.

In the original works~\cite{Friedrich1997,Luck2006,Renner2001} the KM expansion (\ref{eq:km_exp}) was truncated after the second coefficient, which reduces the expansion to an ordinary Fokker-Planck equation (consistent with Kolmogorov-Oboukhov (K62) scaling; see Table~\ref{table:KM}). This truncation is motivated by Pawula's theorem~\cite{Risken}, which states that if an even order KM coefficient $n>2$ is zero then all other coefficients $n>2$ are zero as well.
In this particular case, it can be shown~\cite{Friedrich2011a,Davoudi2000}
that multiscale correlations obey fusion rules (\ref{eq:fusion}). However, the restriction to a Fokker-Planck equation based on the Pawula theorem has proven to be a questionable approximation~\cite{Friedrich2016b,Friedrich2018,Friedrich2017} and higher-order coefficients were found to be small but nonvanishing (see Table \ref{table:KM}). We will show below that the fusion rules are valid even considering the entire KM expansion.
To this end, we cast the solution of Eq.~(\ref{eq:km_exp_trans}) in the form of a Dyson series~\cite{Risken} replacing $r_2=r$ and $r_1=R$,
\begin{eqnarray}\nonumber
 &~&p(v_2,r|v_1,R)\\ \nonumber
 &=&\delta(v_2-v_1) +\int_{r}^{R} \textrm{d} r_1 \hat L_{\rm KM}(v_2, r_1) \delta(v_2-v_1)\\ \nonumber
 &~&+ \int_{r}^{R}\textrm{d} r_1 \int_{r}^{r_1} \textrm{d} r_2
 \hat L_{\rm KM}(v_2,r_1)\hat L_{\rm KM}(v_2,r_2) \delta(v_2-v_1) \\ \nonumber
 &~&+ \ldots
 \\ \nonumber
 &=&\delta(v_2-v_1) +\int_{r}^{R} \textrm{d} r_1 \frac{\hat L(v_2) }{r_1} \delta(v_2-v_1)\\ \nonumber
&~&+ \int_{r}^{R}\textrm{d} r_1 \int_{r}^{r_1} \textrm{d} r_2 \frac{\hat L(v_2)^2}{r_1 r_2}  \delta(v_2-v_1)+ \ldots
 \\ \nonumber
 &=& \delta(v_2-v_1) + \ln \frac{R}{r} \hat L(v_2)
      \delta(v_2-v_1)
 \\ \nonumber
 &~& + \frac{1}{2!} \left(\ln \frac{R}{r}\right)^2 \hat L(v_2)^2 \delta(v_2-v_1)+\ldots \\
 &=& \exp\left[ \ln \frac{R}{r} \hat L(v_2) \right] \delta(v_2-v_1)\;,
\label{eq:dyson}
\end{eqnarray}
where the scale-independent differential operator $\hat L(v_2)$ is defined according to
\begin{equation}
 \hat L(v_2)=\sum_{k=1}^\infty \frac{K_k}{k!}
       \frac{\partial^k}{\partial v_2^k} v_2^k\;.
\end{equation}
Note that the scale ordering problem in the first line of the Dyson series (\ref{eq:dyson}) can be omitted due to the separable form of the KM coefficients (\ref{eq:KM_scaling}).

\begin{table*}[t]
\caption{Phenomenological models of turbulence (we refer the reader
to \cite{Friedrich2016b,Friedrich2018} for further discussions) with scaling exponents $\zeta(n)$
and the corresponding reduced KM coefficients from Eq.~(\ref{eq:KM_scaling}):
Kolmogorov's mean field theory (K41), Kolmogorov-Oboukhov phenomenology (K62), Burgers phenomenology, the $\beta$ model, She-Leveque phenomenology, and Yakhot's model. Note that the K41 theory and the Burgers ramps do not exhibit intermittency corrections. The K62 phenomenology is the only
intermittency model that can be reproduced with just two KM coefficients.
The reduced KM coefficients of the AdS/CFT model~\cite{Eling2015} can only be calculated numerically and have not been included in the table (see~\cite{Friedrich2018,Friedrich2017} for further discussion).\\
$^a\;\mu \approx 0.227$, $^b\;D_F \approx 2.83$, $^c\;$here, $_\nu F_q(a;b;z)$ is the generalized hypergeometric function, $^d\;\beta=0.05$}
\centering
\begin{tabular}{c| c c }
  \toprule[1pt]
   model  & scaling exponent $\zeta(n)$ & reduced KM coefficients $K_n$ \\
   \midrule[0.5pt]
   K41 & $n/3$ & $K_1 = 1/3$ , no higher orders \\
   K62$^a$ & $n/3-\mu n(n-3)/18$ & $\quad K_1 = (3+\mu)/9$,  $K_2 = \mu/9$, no higher orders  \\
   Burgers-ramps & $n$ & $K_1=1$, no higher orders\\
   Burgers-shocks & $1$  & $K_n=1$ \\
   $\beta$-model$^b$ & $\frac{D_F-2}{3}n+(3-D_F)$ & $\quad K_1 = \frac{D_F-2}{3} +(3-D_F)$, $K_n=3-D_F$ for $n>1$\\
   She-Leveque$^c$ &  $\frac{n}{9} + 2\left( 1 - \left( \frac{2}{3}\right)^{n/3} \right)$ &
   $\quad K_n = \frac{1}{9} \left(n_1 F_0(1-n;\ldots;1)+18
   \left(1-\sqrt[3]{\frac{2}{3}}\right)^n\right) $ \\
   Yakhot & $\frac{(1+3\beta)n}{3(1+\beta n)}$  & $K_n =
   \frac{\Gamma[n+1]}{\Gamma[n+1 + \frac{1}{\beta}]}
   \textstyle \left(\Gamma[1+\frac{1}{\beta}] +
   \frac{1}{3 \beta^2} \Gamma[\frac{1}{\beta}] \right)$ \\
  \bottomrule[0.5pt]
\end{tabular}
\label{table:KM}
\end{table*}

We are now in the position to introduce the three-point moments (\ref{eq:fusion}). Due to the ordering $r \le R$, we can take the moments of the two-increment PDF $f_2(v_2,r;v_1,R)=p(v_2,r|v_1,R)f_1(v_1,R)=\left \langle \delta(v_2-\delta_r v)\delta(v_1-\delta_{R} v) \right \rangle$
and obtain
\begin{eqnarray}\nonumber
  &~&\left \langle (\delta_r v)^p (\delta_{R} v)^q \right \rangle\\
   &=& \int \textrm{d}v_2 v_2^p \int \textrm{d}v_1
  v_1^q p(v_2,r|v_1,R)f_1(v_1,R)\;.
 \label{eq:fusion_trans}
\end{eqnarray}
Inserting the Dyson series (\ref{eq:km_exp_trans}) for the transition PDF $p(v_2,r|v_1,R)$ yields
\begin{widetext}
\begin{eqnarray}\nonumber
 \left \langle (\delta_r v)^p (\delta_{R} v)^q \right \rangle
 &=&\left \langle (\delta_{R} v)^{p+q} \right \rangle+
 \ln \frac{R}{r} \sum_{k=1}^{\infty} \frac{K_k}{k!}\int \textrm{d}v_1
 v_1^q \int \textrm{d}v_2\; v_2^p \frac{\partial^k}{\partial v_2^k} v_2^k \delta(v_2-v_1)f_1(v_1,R)\\
 &~& + \frac{1}{2!} \left(\ln \frac{R}{r}\right)^2 \sum_{k=1}^{\infty}\sum_{l=1}^{\infty} \frac{K_k K_l}{k!l!}
 \int \textrm{d}v_1
 v_1^q \int \textrm{d}v_2\, v_2^p \frac{\partial^k}{\partial v_2^k} v_2^k \frac{\partial^{l}}{\partial v_2^{l}} v_2^{l}
 \delta(v_2-v_1)f_1(v_1,R)+\ldots
\end{eqnarray}
\end{widetext}
Partial integrations with respect to $v_2$ in the second and third term yields
\begin{widetext}
  \begin{eqnarray}\nonumber
   \left \langle (\delta_r v)^p (\delta_{R} v)^q \right \rangle
 &=& \left \langle (\delta_{R} v)^{p+q} \right \rangle+
 \ln \frac{R}{r} \sum_{k=1}^{p} \frac{(-1)^k K_k p!}{k! (p-k)!}\int \textrm{d}v_1
 v_1^q \int \textrm{d}v_2\, v_2^p \delta(v_2-v_1)f_1(v_1,R) \\ \nonumber
 &~&  +\frac{1}{2!} \left(\ln \frac{R}{r}\right)^2 \sum_{k=1}^{p} \frac{(-1)^k K_k p!}{k! (p-k)!}
 \sum_{l=1}^{p} \frac{(-1)^{l} K_l p!}{l! (p-l)!}
 \int \textrm{d}v_1
 v_1^q \int \textrm{d}v_2\, v_2^p \delta(v_2-v_1)f_1(v_1,R)+\ldots\\ \nonumber
 &=&  \left[1+ \ln \frac{R}{r} \sum_{k=1}^{p} (-1)^k K_k{ p \choose k}  +\frac{1}{2!} \left(\ln \frac{R}{r}\right)^2 \sum_{k=1}^{p} (-1)^k K_k{ p \choose k}
 \sum_{l=1}^{p} (-1)^{l} K_l{ p \choose l} +\ldots\right ]\left \langle (\delta_{R}v)^{p+q} \right \rangle \\ \nonumber
 &=& \left[ 1 - \ln \frac{R}{r} \zeta(p)
 +\frac{1}{2!} \left(\ln \frac{R}{r}\right)^2 \zeta(p)^2 +\ldots \right] \left \langle (\delta_{R} v)^{p+q} \right \rangle=
 \exp \left[ -\zeta(p)\ln \frac{R}{r} \right] \left \langle (\delta_{R} v)^{p+q} \right \rangle
 \\
 &=& \exp \left[ \zeta(p)\ln \frac{r}{R} \right] \left \langle (\delta_{R} v)^{p+q} \right \rangle
 = \frac{r^{\zeta(p)}}{R^{\zeta(p)}} \left \langle (\delta_{R} v)^{p+q} \right \rangle=
\frac{\langle (\delta_r v)^p\rangle}{\langle (\delta_{R} v)^p\rangle} \langle (\delta_{R} v)^{p+q}\rangle \;.
\label{eq:fusion_markov}
\end{eqnarray}
\end{widetext}
Here we made use of the relation (\ref{eq:zeta}) and inserted $\langle (\delta_r v)^p \rangle \sim r^{\zeta(p)}$ in the last step. In other words,
the operator-product expansion can be conceived as a Markov process of velocity increments in scale, a direct consequence of the multiplicative process (\ref{eq:multiplier}) and its uncorrelated multipliers.
Empirical evidence suggests that the multiplicative uncorrelated fusion-rule prediction (\ref{eq:fusion_markov}) breaks down in the limit of $r \rightarrow R$. In terms of the Markov property (\ref{eq:markov}), such a violation can be explained by the existence of nontrivial correlations in the energy transfer for not-too-separated scales.

In conclusion, the application of the fusion rules (\ref{eq:fusion}) necessarily entails two aspects:
 (i) the validity of the Markov property of velocity increments in scale (\ref{eq:markov}), which implies that the KM expansion for the transition PDF (\ref{eq:km_exp_trans}) conforms with the KM expansion for the one-increment PDF (\ref{eq:km_exp}), and
  (ii) the specific form of the KM coefficients (\ref{eq:KM_scaling}) which was chosen in a way to ensure the existence of scaling solutions $\langle \delta_r v^n \rangle \sim r^{\zeta(n)}$.

For the sake of completeness, we want to end this section with a generalization of fusion rules (\ref{eq:fusion}) to $n$-increment statistics [($n+1$)-point statistics in terms of ordinary moments]. The procedure follows along the same lines as the derivation of the fusion rules from the KM expansions of the Markov process (\ref{eq:fusion_markov}) and is explained in Appendix \ref{app:gen_fusion}. We obtain
\begin{eqnarray}
  \lefteqn{\left \langle (\delta_{r_n}v)^{p_n} \cdots (\delta_{r_2}v)^{p_2}(\delta_{r_1}v)^{p_1} \right \rangle}\\ \nonumber
  &=&
  \int \textrm{d}v_n \cdots \textrm{d}v_2\; \textrm{d}v_1\;  v_n^{p_n}\ldots v_2^{p_2}\; v_1^{p_1}\; f_n(v_n,r_n;\cdots v_1,r_1)\\ \nonumber
  &=& \prod_{i=1}^{n-1} \frac{\left \langle (\delta_{r_{i+1}}v)^{\sum_{k=1}^{i} p_{n+1-k}} \right \rangle}
  {\left \langle (\delta_{r_i}v)^{\sum_{k=1}^{i}p_{n+1-k}} \right \rangle}
  \left \langle (\delta_{r_1}v)^{\sum_{k=1}^{n}p_{n+1-k}} \right \rangle,
  \label{eq:gen_fusion}
\end{eqnarray}
where $f_n$ is the $n$-increment PDF (\ref{eq:n_inc_pdf}). These generalized fusion rules imply a reduction of am ($n+1$)-point statistical quantity to a two-point quantity.

\section{Application to Burgers turbulence}
\label{sec:burgers_app}
In contrast to the dissipation anomaly that arises in the MF description (Sec.~\ref{sec:fusion}), the dissipation anomaly that arises in the multiscale description of Burgers turbulence bears a clear physical meaning: Due to the absence of nonlocal pressure contributions, singular structures consist of localized shocks whose widths are determined by the viscosity $\nu$. For example, consider the single shock solution of Eq.~(\ref{eq:burgers}),
\begin{equation}
 v(x,t) = 1- \tanh \left(\frac{x-x_c-t}{2 \nu}\right)\;,
 \label{eq:shock}
\end{equation}
where the width of the shock is inversely proportional to $\nu$. It can be readily seen that the averaged local energy dissipation rate $\langle \varepsilon \rangle$, where
\begin{equation}
 \varepsilon(x)  =
 2\nu \left(\frac{\partial v(x)}{\partial x}\right)^2
 \label{eq:varepsilon}
\end{equation}
is independent of the viscosity $\nu$.
In the following, we will further discuss multiscale properties of the Burgers equations, including inertial-viscous cases such as the ones described by the correlations (\ref{eq:diss_anomaly_mf}).\\

\subsection{Dissipation anomaly in a multi-increment PDF hierarchy in Burgers turbulence}\label{sec:burgers}
We consider the Burgers equation
\begin{equation}
 \frac{\partial }{\partial t} v(x,t) + v(x,t) \frac{\partial}{\partial x} v(x,t)
 = \nu \frac{\partial^2}{\partial x^2}v(x,t) +F(x,t),
 \label{eq:burgers}
\end{equation}
with a white noise in time Gaussian forcing $F(x,t)$ defined by the second order moment
\begin{equation}
 \langle F(x,t) F(x',t) \rangle = \chi(x-x') \delta(t-t')\;,
 \label{eq:forcing_corr}
\end{equation}
where $\chi(x-x')$ is the spatial correlation function, assumed to be concentrated around a characteristic scale $|x-x'| \sim l_f$.
The evolution equation for the velocity increment $\delta_r v(x,t)$ is
\begin{widetext}
\begin{eqnarray}
 \frac{\partial  \delta_r v(x,t)}{\partial t} + v(x,t) \frac{\partial \delta_r v(x,t)}{\partial x}  + \delta_r v(x,t) \frac{\partial \delta_r v(x,t)}{\partial r}
 = \nu \frac{\partial^2 \delta_r v(x,t)}{\partial x^2}
 +F(x+r,t)-F(x,t)\;.
 \label{eq:burgers_inc}
\end{eqnarray}
\end{widetext}
The temporal evolution of the one-increment PDF (\ref{eq:inc}) is derived in Appendix \ref{app:1} according to
\begin{widetext}
\begin{eqnarray}\nonumber
&~& \frac{\partial}{\partial t} f_1(v_1,r_1,t)+ v_1 \frac{\partial}{\partial r_1}
f_1(v_1,r_1,t)+2\int_{- \infty}^{v_1} \textrm{d} v_1' \frac{\partial}{\partial r_1} f_1(v_1',r_1,t)\\
 &=&
 -\nu \frac{\partial}{\partial v_1} \int \textrm{d} r_2 \left[ \delta(r_2-r_1) -
 \delta(r_2) \right] \frac{\partial^2}{\partial r_2^2}
 \int \textrm{d} v_2 v_2 f_2(v_2,r_2;v_1,r_1,t)
  +[\chi(0)-\chi(r_1)] \frac{\partial^2 }{\partial v_1^2}f_1(v_1,r_1,t)\;.
  \label{eq:evo1}
\end{eqnarray}
\end{widetext}
Due to the viscous coupling to the two-increment PDF, we have a hierarchy formally similar to the Bogoliubov-Born-Green-Kirkwood-Yvon statistical physics case~\cite{Polyakov1995a,Weinan1999}.

It is useful to  reformulate the dissipative terms in order to introduce the local energy dissipation rate (\ref{eq:varepsilon}).
First, we assume the stationarity of the velocity increment statistics, i.e.,
$\frac{\partial}{\partial t}f_1(v_1,r_1,t)$=0. Second, as shown in Appendix \ref{app:2}, the unclosed viscous term in Eq.~(\ref{eq:evo1}) can be rewritten  in terms of the joint velocity gradient and
velocity increment statistics as
\begin{eqnarray}\nonumber
 &~&v_1 \frac{\partial}{\partial r_1}
f_1(v_1,r_1)
 =  -2\int_{- \infty}^{v_1} \textrm{d} v_1' \frac{\partial}{\partial r_1} f_1(v_1',r_1)
 \\ \nonumber
   &~& -\frac{\partial^2 }{\partial v_1^2}\Bigg[ \left \langle \frac{\varepsilon(x)}{2}
   [\delta(v_1-\delta_{r_1} v(x))+ \delta(v_1+\delta_{-r_1} v(x))] \right \rangle \\
  &~&  +[\chi(0)-\chi(r_1)]f_1(v_1,r_1) \Bigg] + 2\nu \frac{\partial^2}{\partial r_1^2} f_1(v_1,r_1)\;.
   \label{eq:evo1_eps}
\end{eqnarray}
From this expression, the existence of the dissipative anomaly becomes
more apparent than in Eq.~(\ref{eq:evo1}) due to the nonvanishing local energy dissipation rate in the limit $\nu \rightarrow 0$.
Taking the moments of Eq.~(\ref{eq:evo1_eps}) and dropping the index of $r_1$ yields
\begin{eqnarray}\nonumber
  &~&\left(1-\frac{2}{n}\right) \frac{\partial}{\partial r} \left\langle [\delta_r v(x)]^n \right \rangle\\ \nonumber
 &=& 2\nu \frac{\partial^2}{\partial r^2}\langle [\delta_r v(x)]^{n-1}
 \rangle \\ \nonumber
 &~&- \frac{(n-1)(n-2)}{2} \left \langle  \varepsilon(x)
 \{[\delta_r v(x)]^{n-3}+(-\delta_{-r}v(x))^{n-3}\}
\right \rangle \\
 &~&+ (n-1)(n-2) [ \chi(0) -\chi(r)] \left \langle [\delta_r v(x)]^{n-3} \right \rangle\;.
 \label{eq:moment_evo}
\end{eqnarray}
For $n=3$, we recover the equivalent of Kolmogorov's $4/5$ law for Burgers turbulence
\begin{eqnarray}\nonumber
  \frac{1}{3} \frac{\partial}{\partial r} \left\langle [\delta_r v(x)]^3 \right \rangle
 &=& -2 \langle \varepsilon \rangle + 2\nu \frac{\partial^2}{\partial r^2}\langle [\delta_r v(x)]^2\rangle \\ \
 &~&+ 2[\chi(0) -\chi(r)] \;,
 \label{eq:4/5_start}
\end{eqnarray}
which reduces to $\left\langle (\delta_r v)^3 \right \rangle = -6 \langle \varepsilon \rangle r$ in the inertial range.

\begin{figure}[ht!]
  \includegraphics[width=0.49 \textwidth]{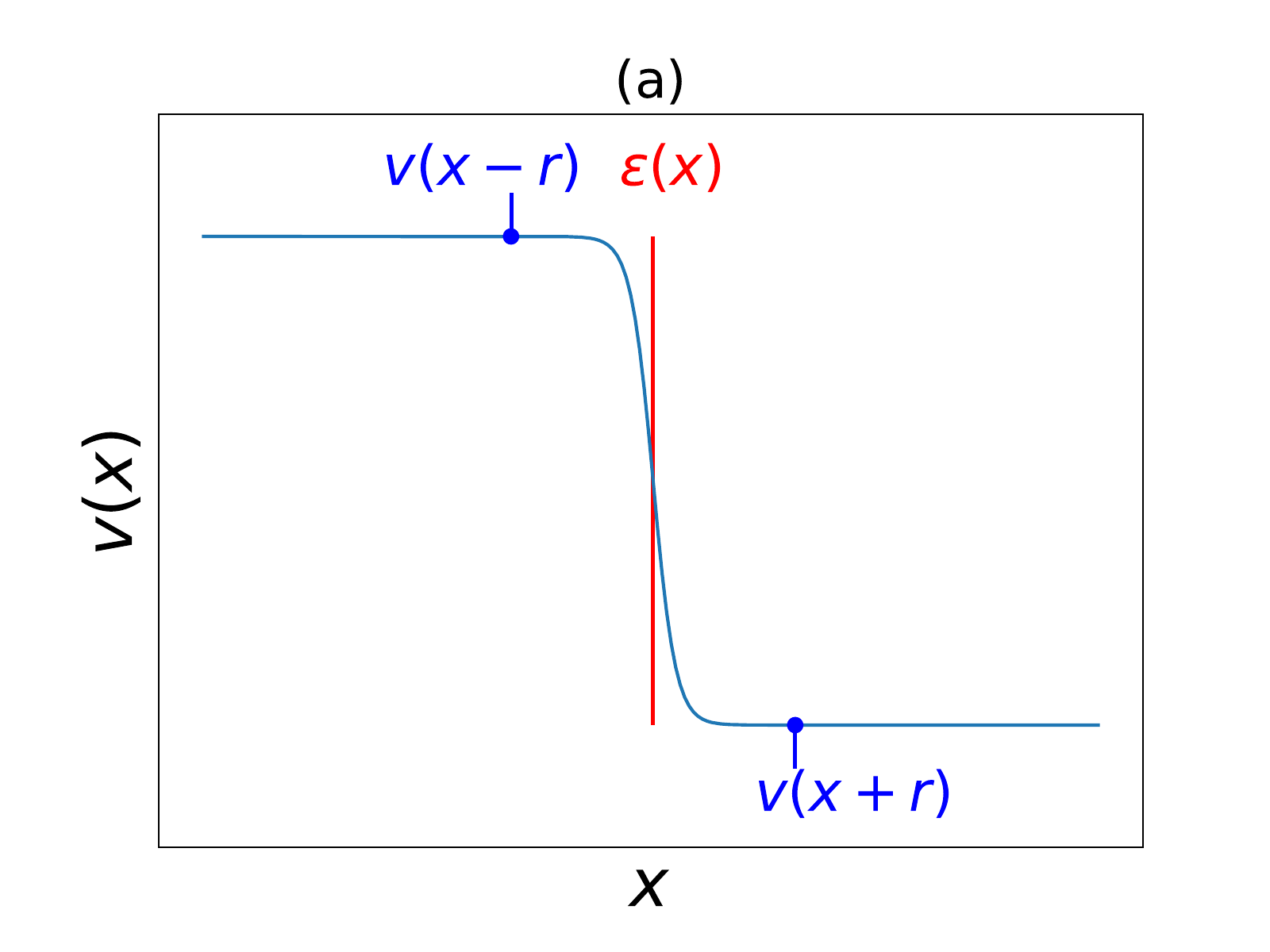}
  \includegraphics[width=0.49 \textwidth]{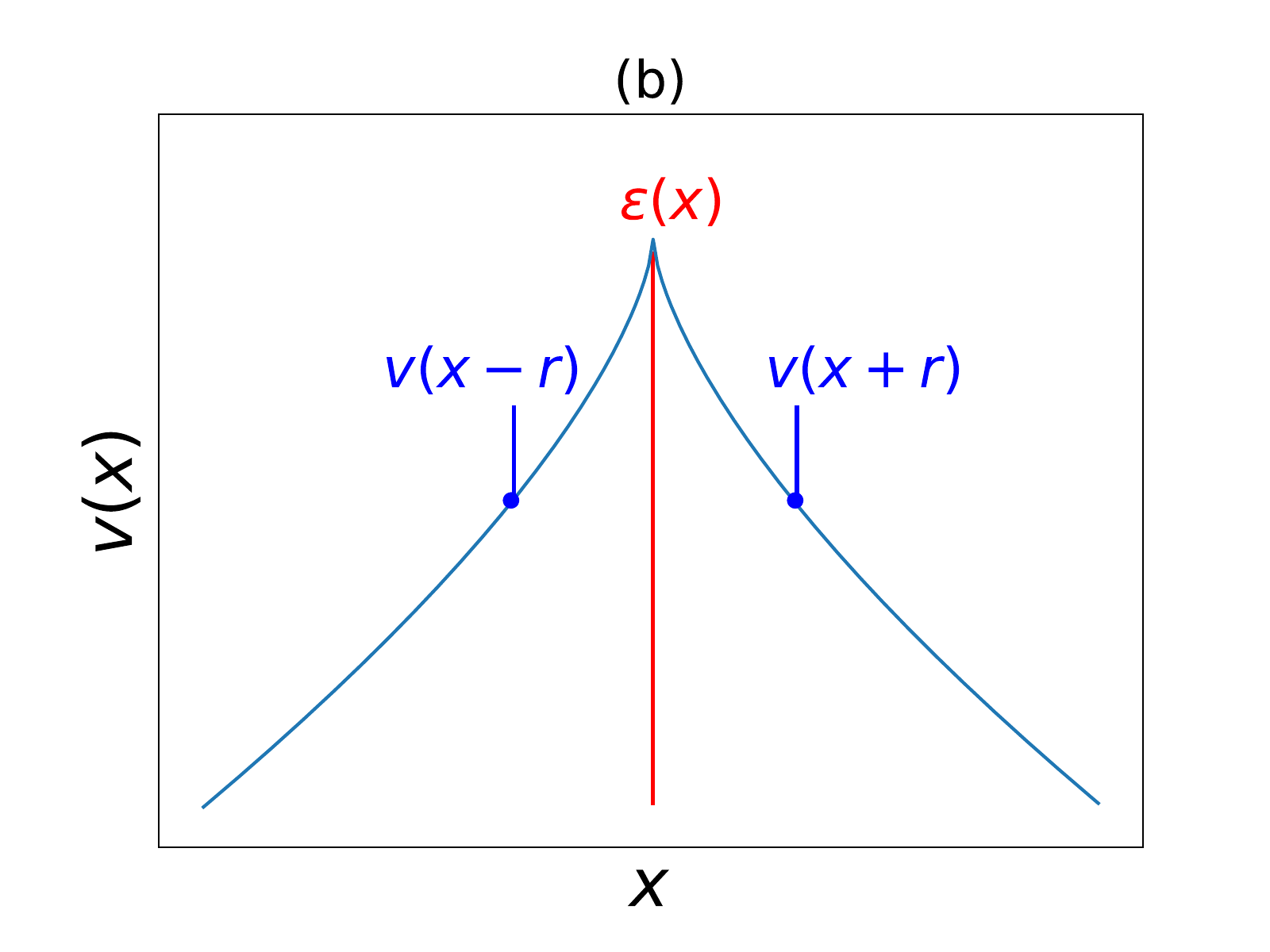}
  \caption{(a) Schematic depiction of a shock in Burgers turbulence. The local energy dissipation rate is peaked at the center of the shock $\varepsilon(x)$. Depending on the strength of the shock, the velocity field at $v(x-r)$ and at $v(x+r)$ possesses the symmetry
  $v(x-r)=-v(x+r)$, which leads to Burgers scaling~(\ref{eq:anomaly_scaling1}).
  (b) In the case of cusplike structures, $\varepsilon(x)$ is still peaked in the center of the cusp. The symmetry of the cusps, however, leads to the vanishing of the dissipation anomaly in Eq.~(\ref{eq:moment_evo}).}
\label{fig:anomaly}
\end{figure}

In the general case, i.e., for $n \neq 3$,
we start by discarding the forcing contribution in the inertial range $\eta \ll r \ll L$ in assuming that $\chi(r)$ decreases sufficiently fast for increasing $r$. Moreover, in the limit of high Reynolds numbers, i.e., $\nu \rightarrow 0$, the smooth subleading viscous term
$2\nu \frac{\partial^2}{\partial x^2}\langle (\delta_r v)^{n-1} \rangle$ can be neglected. Hence, in the inertial range where
 $\left \langle (\delta_r v)^{n-3} \right \rangle$ should admit scaling, we obtain
\begin{equation}
 \left \langle  \varepsilon(x) [\delta_r v(x)]^n\right \rangle
 \sim  |r|^{\zeta(n+3)-1} ,
 \label{eq:burgers_mult}
\end{equation}
which agrees with the first result (\ref{eq:diss_fusion}) of the MF model.

Hence, the prediction made by the MF model (\ref{eq:diss_fusion})
becomes exact for the case of Burgers turbulence. It must be stressed that  (\ref{eq:moment_evo}) does not further specify the scaling exponent $\zeta(n)$. It is well known that in order to go beyond it, we need some heuristic arguments  about the dissipative term based on the geometrical structures of the flow. In high-Reynolds-number Burgers turbulence, we are faced with shocklike structures similar to the one in Fig.~\ref{fig:anomaly}(a). In this case, the local energy dissipation rate is peaked at the center of the shock and the velocities $v(x+r)$ and $v(x-r)$ are arranged antisymmetrically around $v(x)=0$. In the limit of small viscosities and for small $r$, $v(x \pm r)$ possesses a negligible dependence on $r$ and we obtain
\begin{eqnarray}\nonumber
  &~&\big \langle  \varepsilon(x)
 \{[\underbrace{\delta_r v(x)}_{=v(x+r)}]^{n-3}+[\underbrace{-\delta_{-r}v(x)}_{=-v(x-r)}]^{n-3}\}
\big \rangle \\
&\sim& \langle v^2 \rangle^{(n-3)/2} \langle \varepsilon \rangle \sim r^{\zeta(n)-1} \quad \rightarrow \quad \zeta(n) =1\;.
  \label{eq:anomaly_scaling1}
\end{eqnarray}
This is exactly the celebrated Burgers shock scaling from Table \ref{table:KM}. It is important to stress that there exists a series of rigorous and quasirigorous results on the PDF of the gradient statistics in Burgers equations \cite{Chekhlov1995,Yakhot1996,Polyakov1995a,Boldyrev1997,Boldyrev1998,Kraichnan1957,Balkovsky1997,Chernykh:2001,EWeinan1997,Weinan1999,Eijnden2000}. It is generally believed that it develops power-law tails in the inviscid limits. In our derivation, we do not pretend to control leading and sub-leading contributions in the zero-viscosity case. Our treatment is limited to estimate the regime of high  but finite Reynolds number.

The influence of smooth velocity field structures can be seen as follows:
Consider Eq.~(\ref{eq:moment_evo}) for small $r$, in which case we can neglect the nonlinear and forcing contributions.
\begin{eqnarray} \nonumber
  &~&2\nu \frac{\partial^2}{\partial r^2}\langle [\delta_r v(x)]^{n-1}
  \rangle \\ \nonumber
  &\approx& \frac{(n-1)(n-2)}{2} \left \langle  \varepsilon(x)
  \{[\delta_r v(x)]^{n-3}+[-\delta_{-r}v(x)]^{n-3}\}
  \right \rangle \\
  &\approx&  2 \nu (n-1)(n-2) \left \langle  \left(\frac{\partial v(x)}{\partial x} \right)^{n-1}r^{n-3}
  \right \rangle\;,
  \label{eq:smooth}
\end{eqnarray}
where we performed a Taylor expansion $\delta_r v(x)= v(x+r)-v(x) \approx \frac{\partial v(x)}{\partial x}r$ inside the ensemble average on the right-hand side
and replaced the local energy dissipation rate $\varepsilon(x)$ with its definition (\ref{eq:varepsilon}). Integrating Eq.~(\ref{eq:smooth}) and reinserting the definition of the local energy dissipation rate (\ref{eq:varepsilon}) yields
\begin{equation}
  \langle [\delta_r v(x)]^{n}
  \rangle = 2^{-n/2-1/2}\frac{\langle \varepsilon^{n/2} \rangle}{\nu^{n/2-1/2}} r^n .
  \label{eq:burgers_scaling_smooth}
\end{equation}
Obviously, this result bears the signature of smooth ramp like velocity field contributions $v(x)$ in between shocks and is the leading term for $n<1$. Hence, by including the heuristic result (\ref{eq:anomaly_scaling1}), we obtain the well-known Burgers scaling
\begin{equation}
   \langle |\delta _r v|^n \rangle \sim
  \begin{cases}
  \, r^n & \textrm{ for } n < 1  \\
  \, r & \textrm{ for } n \geq 1\,.
  \end{cases}
  \label{eq:burgers_scaling}
\end{equation}

In order to understand the importance of the exact shape of the singularity, it is instructive to consider the case of the Burgers equation with an additional nonlocality~\cite{Zikanov1997,Friedrich2016b}.
\begin{equation}
 \frac{\partial }{\partial t} v(x,t) + w(x,t) \frac{\partial}{\partial x} v(x,t)
 = \nu \frac{\partial^2}{\partial x^2}v(x,t) +F(x,t),
 \label{eq:gen_burgers}
\end{equation}
where the convective velocity field is given by
\begin{equation}
w(x,t)= \alpha v(x,t) +(1-\alpha) \textrm{P.V.}\int \textrm{d} x' \frac{v(x',t)}{x-x'}~,
\end{equation}
where P.V. denotes principal value. Here $\alpha=1$ corresponds to the case of Burgers turbulence,
whereas $\alpha=0$ corresponds
to the purely nonlocal case that exhibits self-similar behavior~\cite{Zikanov1997}. In the latter case, the velocity field is dominated by cusplike structures similar to the one depicted in Fig.~\ref{fig:anomaly}(b). Consequently, the velocity field possesses the symmetry $v(x-r)=v(x+r)$ leading to the vanishing of the dissipative term
$\left \langle  \varepsilon(x)
\{[\delta_r v(x)]^{n-3}+[-\delta_{-r}v(x)]^{n-3}\}
\right \rangle$ for even $n$. Furthermore, the nonlinear terms in the PDF hierarchy are changed due to the presence of the nonlocality in the generalized Burgers equation (\ref{eq:gen_burgers}) and are necessarily unclosed~\cite{Friedrich2017}.
Accordingly, the nonlinear terms in the purely nonlocal case are balanced by the forcing terms. Depending on the properties of the forcing correlation function this scaling can be associated with the results of the renormalization group (see~\cite{McCo} for further references) and necessarily implies non-intermittent scaling.

Another important case of Eq.~(\ref{eq:moment_evo}) is when the local dissipation rate and the velocity increment are statistically independent
\begin{eqnarray}\nonumber
  &~& \left \langle  \varepsilon(x)
  \{[\delta_r v(x)]^{n-3}+[-\delta_{-r}v(x)]^{n-3}\}
 \right \rangle \\ \nonumber
     &~& = \underbrace{  \left \langle
     \{[\delta_r v(x)]^{n-3}+[-\delta_{-r}v(x)]^{n-3}\}
    \right \rangle}_{\sim r_1^{\zeta(n-3)}}
       \left \langle \varepsilon(x)\right \rangle\\ \nonumber
        &\sim& r^{\zeta(n)-1}
     \rightarrow \;\zeta(n)-1= \zeta(n-3) \;\rightarrow
     \;  \zeta(n) =n/3.
     \\
   \label{eq:anomaly_scaling2}
 \end{eqnarray}
which necessarily implies Kolmogorov (K41) scaling. The case of  Burgers scaling (\ref{eq:anomaly_scaling1}) must be considered as the  opposite case:  The energy dissipation rate is fully correlated with the velocity increment, leading to strong intermittency. Furthermore, it has been shown  that the intermediate case $0<\alpha<1$ in Eq.~(\ref{eq:gen_burgers}) shares many resemblances with the original Navier-Stokes equation \cite{Zikanov1997,Friedrich2016b}. Accordingly, the pressure must have a regularizing effect on the velocity field structures that enter the dissipation anomaly.

In the following section, we will evaluate both the fusion rules from Sec.~\ref{sec:markov_fusion} and the multifractal prediction from direct numerical simulations of Burgers turbulence.

\subsection{Direct numerical simulations of Burgers turbulence}
\label{sec:dns}
In order to validate the theoretical considerations of the previous sections, we performed direct numerical simulations (DNSs) of the stochastically driven Burgers equation (\ref{eq:burgers}). The numerical setup consists of a second order Adams-Bashforth explicit solver paired with an Euler-Maruyama step to account for the large-scale Gaussian random forcing. We also consider the variable transformation $\hat v'_k(t) = \exp ( -\nu \, k^2 \,dt ) \hat v_k(t)$, which implies the exact integration of the viscous term. It relaxes the restriction on the time step by the diffusive term and significantly improves the convergence for large wave numbers.
The spatial correlation function of the forcing (\ref{eq:forcing_corr}) follows a power law proportional to $k^{-2}$ in Fourier space and has a cutoff at $k_F = 5$. Table \ref{tab:parameters} contains a list of the characteristic parameters in use for the simulations
presented in Figs.~\ref{fig:fusion}--\ref{fig:multifractal}. The resolution was fixed such  that
$\eta / \textrm{d}x \approx 6$ at the highest Reynolds number. To improve the statistics we averaged over 200 independent runs.

\begin{table}
\caption{ Characteristic parameters of the
numerical simulations: root-mean-square velocity $v_{\rm rms}= \sqrt{ \langle v^2 \rangle}$,
viscosity $\nu$,
averaged rate of local energy dissipation
$\langle \varepsilon \rangle= 2
\nu \left\langle \left(\frac{\partial v}{\partial x} \right)^2 \right\rangle$,
grid spacing $dx$, timestep $dt$,
dissipation length $\eta=\left(\frac{\nu^3}{\langle \varepsilon \rangle}
\right)^{1/4}$,
Taylor length $\lambda = v_{\rm rms} \sqrt{\frac{2\nu}{\langle \varepsilon \rangle }}$,
Taylor-Reynolds number $\textrm{Re}_{\lambda}=
\frac{v_{\rm rms} \lambda}{\nu}$,
integral length scale $L= \frac{E_{\rm kin}^{3/2}}{\langle \varepsilon \rangle}$,
kinetic energy $E_{\rm kin}=\frac{1}{2}v_{\rm rms}^2$, large-eddy turnover time
$T_L=\frac{L}{v_{\rm rms}}$, number of grid points $N$, and maximum force wavenumber of the power-law forcing $k_F$. The physical domain size is $2\pi$.}
\centering
\begin{tabular}{c c c c}
  \toprule[1pt]
  \toprule[1pt]
  Parameter \qquad  & Run 1 \qquad & Run 2 \qquad & Run 3 \qquad \\
  \midrule[0.5pt]
  $u_{\rm rms}$ &  1.16 & 1.16    & 1.15\\
  $\nu$ & $3.6 \times 10^{-4}$ &    $1.2 \times 10^{-3}$  & $6.8 \times 10^{-3}$\\
  Re  & 1800   & 550 & 90\\
  Re$_{\lambda}$ & 100   & 56 & 23\\
  $\langle \varepsilon \rangle$ & 1 & 1  & 1\\
  $dt$ & $1.53 \times 10^{-5}$    & $1.53 \times 10^{-5}$  & $1.53 \cdot 10^{-5}$\\
  $dx$ &   $3.83 \times 10^{-4}$   & $3.83 \times 10^{-4}$  & $3.83 \cdot 10^{-4}$\\
  $\eta$ &   $2.61 \times 10^{-3}$   & $6.31 \times 10^{-3}$  & $2.37 \times 10^{-2}$\\
  $\lambda$ & 0.031 & 0.056 & 0.134\\
  $L$ & 1.564 & 1.555  & 1.526\\
  $T$ in $T_L$ & 760  & 762 & 772\\
  $N$ & $2^{14}$ & $2^{14}$ & $2^{14}$\\
  $k_F$ & 5 & 5 & 5\\
  \bottomrule[0.5pt]
  \bottomrule[0.5pt]
\end{tabular}
\label{tab:parameters}
\end{table}
\subsubsection{Evaluation of inertial-inertial fusion rules from DNS of Burgers turbulence}
First, we investigate the validity of the fusion rules (\ref{eq:fusion})
for the Burgers equation. To this end, we consider the quantity
\begin{equation}
  F_{p,q}(r,R)= \langle (\delta_r v)^p (\delta_{R} v)^q \rangle\;.
  \label{eq:f}
\end{equation}
The application of the fusion rules (\ref{eq:fusion}) in conjunction with the Burgers scaling (\ref{eq:burgers_scaling}) yields three different possible scaling properties, depending on the order of the moments $p$ and $q$.
If both increments are dominated by the shock we have case I, if both are dominated by the smooth ramps we have the case II, and if the small scale is smooth and the large scale is dominated by the shock we have case III.
The scaling prediction in the plane $(p,q)$  is summarized in Fig.~\ref{fig:fusion_rls_pred} and as follows:
\begin{figure}[t]
  \centering
  \includegraphics[width=0.46 \textwidth]{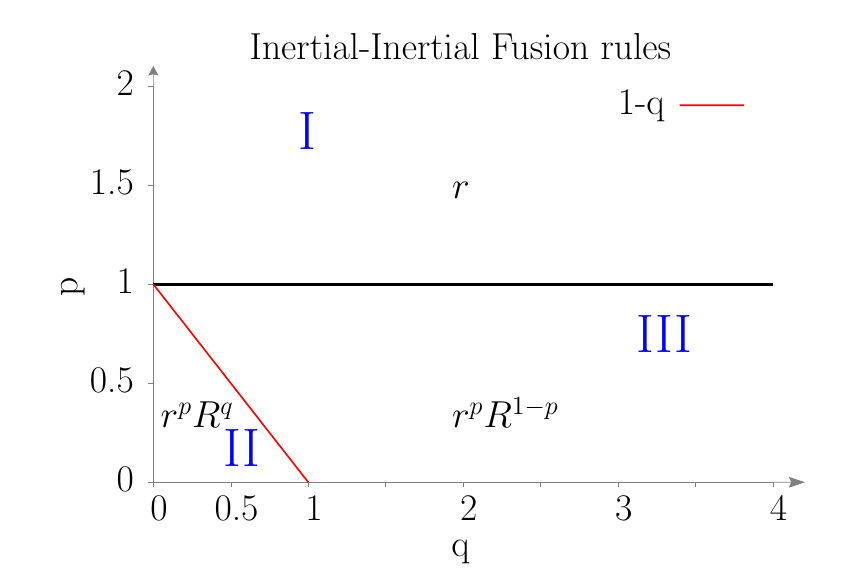}
  \caption{Parametric space of the exponents of Eq.~(\ref{eq:f}) and the
corresponding fusion rule prediction. Here I--III correspond to the three regions that emerge from the bifractal Burgers scaling (\ref{eq:burgers_scaling}). }
\label{fig:fusion_rls_pred}
\end{figure}
\begin{equation}
  F_{p,q}(r,R) \sim
  \begin{cases}
  \, r & \text{for $p > 1$}\, \\
  \, r^p \, R^q & \text{for $p < 1$ \& $p+q \leq 1$}\, \\
  \, r^p \, R^{1-p} & \text{for $p < 1$ \& $p+q \geq 1$}\,.
  \end{cases}
  \label{eq:inert_inert_scaling}
\end{equation}

\begin{figure}[t!]
  \includegraphics[width=0.49 \textwidth]{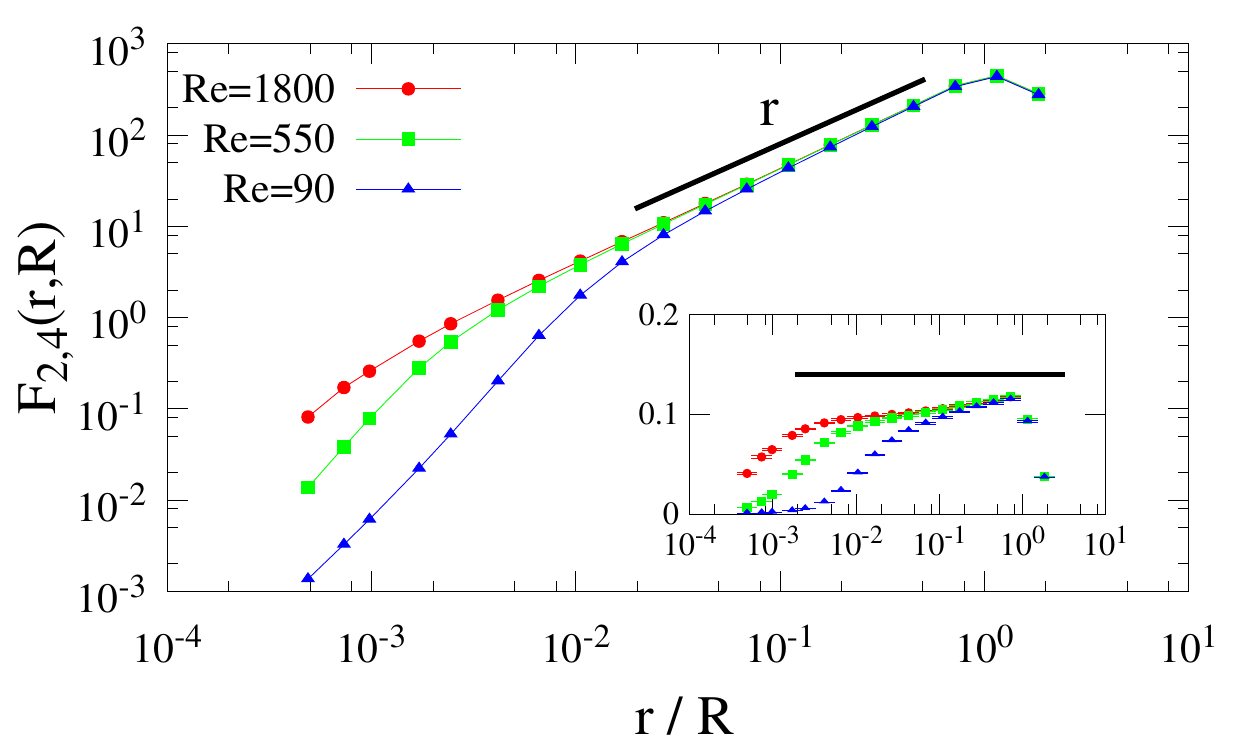}
  \includegraphics[width=0.49 \textwidth]{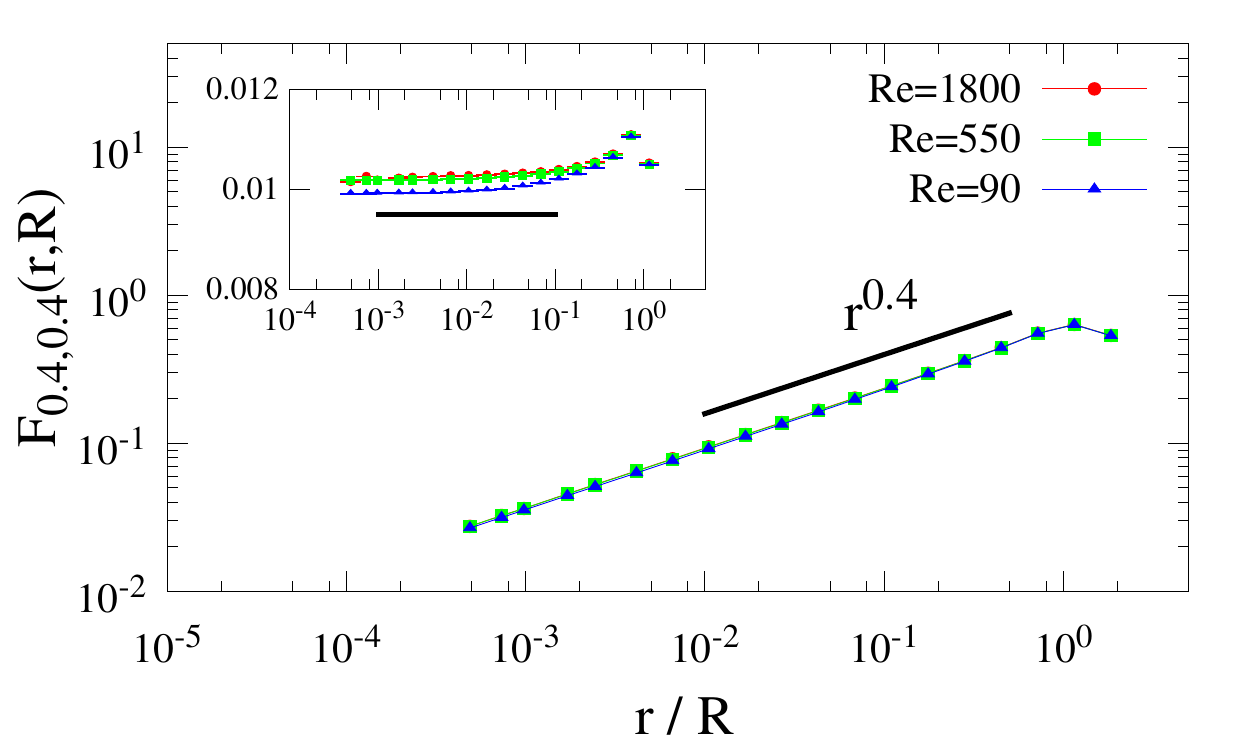}
  \includegraphics[width=0.49 \textwidth]{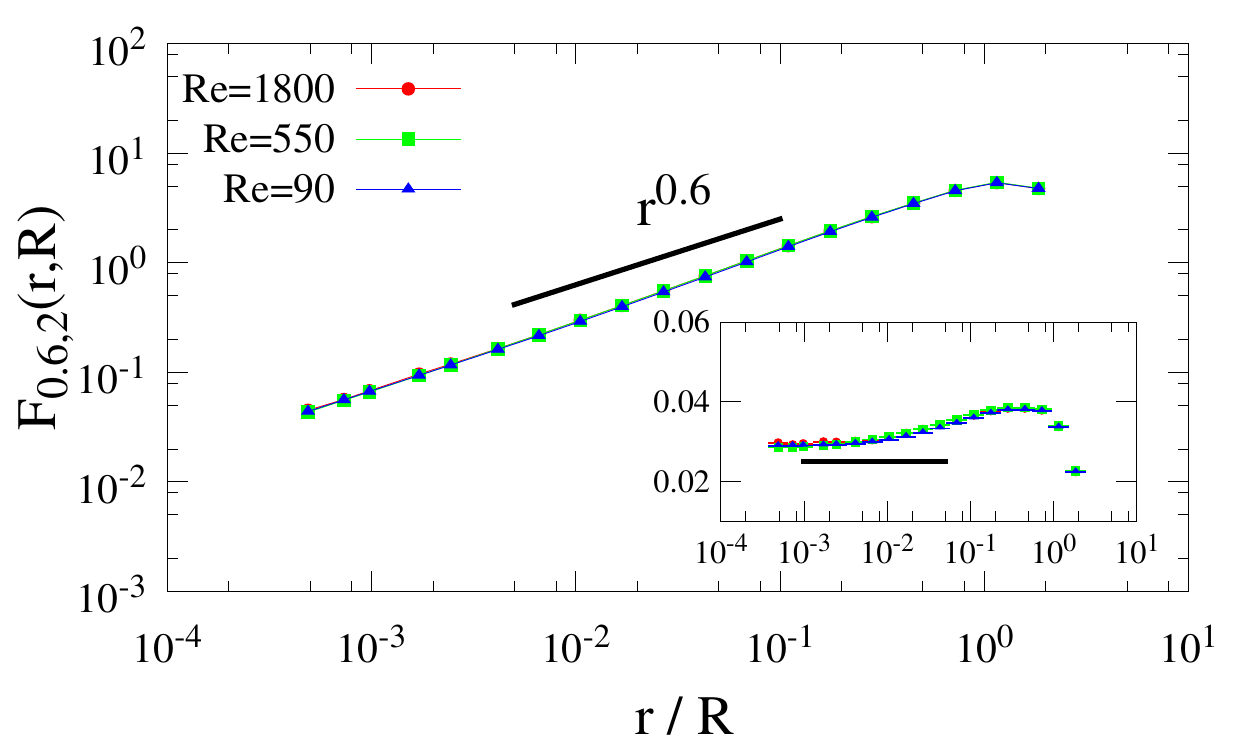}
  \caption{Examination of the inertial-inertial fusion rules (\ref{eq:inert_inert_scaling}) for runs 1--3 via the quantity $F_{p,q}(r,R)$ defined in Eq.~(\ref{eq:f}) as a function of $r/R$ for different values of $p$ and $q$ and for different Reynolds numbers: (a) $p=2$ and $q=4$ (region I). (b) $p=q=0.4$ (region II). (c) $p=0.6$ and  $q=2$ (region III). The black line corresponds to the fusion-rule prediction in Burgers turbulence summarized in the three regions  I--III of Fig.~\ref{fig:fusion_rls_pred}. The inset depicts $F_{p,q}(r,R)$  divided by this prediction. }
\label{fig:fusion}
\end{figure}

In the following, we set the large scale $R$ equal to $\pi/2$ and vary the small scale $r$. The large scale $R$ is fixed so that $R/\eta \approx 600$ for $\textrm{Re}=1800$, $R/\eta\approx 250$ for $\textrm{Re}=550$, and $R/\eta \approx 70$ for $\textrm{Re}=90$. We have also tested the opposite scenario by fixing the small scale $r$ to $\alpha \eta$ with $\alpha \in [2,10]$,
which yielded similar results that will therefore not be shown here. Figure \ref{fig:fusion} depicts $F_{p,q}(r,R)$ for three values in the three regions of Fig.~\ref{fig:fusion_rls_pred}: ($p=2,q=4$) [Fig.~\ref{fig:fusion}(a), region I], ($p=q=0.4$) [Fig.~\ref{fig:fusion}(b), region II], and ($p=0.6,q=2$) [Fig.~\ref{fig:fusion}(c), region III]. As one can see, all three cases agree fairly well with
the theoretical predictions (black lines with the corresponding predicted scaling). This becomes even more apparent from the insets in Fig.~\ref{fig:fusion}, which shows
$F_{p,q}(r,R)$ compensated by the corresponding prediction. We observe constant ($r$-independent) regions over a few decades of $r / R$. However, as $r$ approaches larger values and tends towards $R$, the compensated function becomes $r$-dependent, which indicates a breakdown of the fusion rules for small-scale separations. As discussed in Sec.~\ref{sec:markov_fusion}, the breakdown of the fusion rules for small-scale separations can also be interpreted in terms of the violation of the
Markov property (\ref{eq:markov}).
In the following section, we will consider the special case of $F(r,R)$ for $r \rightarrow \eta$ to check the viscous-inertial scaling.

\subsubsection{Evaluation of the viscous-inertial fusion rules prediction from DNS of Burgers turbulence}
\begin{figure}[h]
  \centering
  \includegraphics[width=0.46 \textwidth]{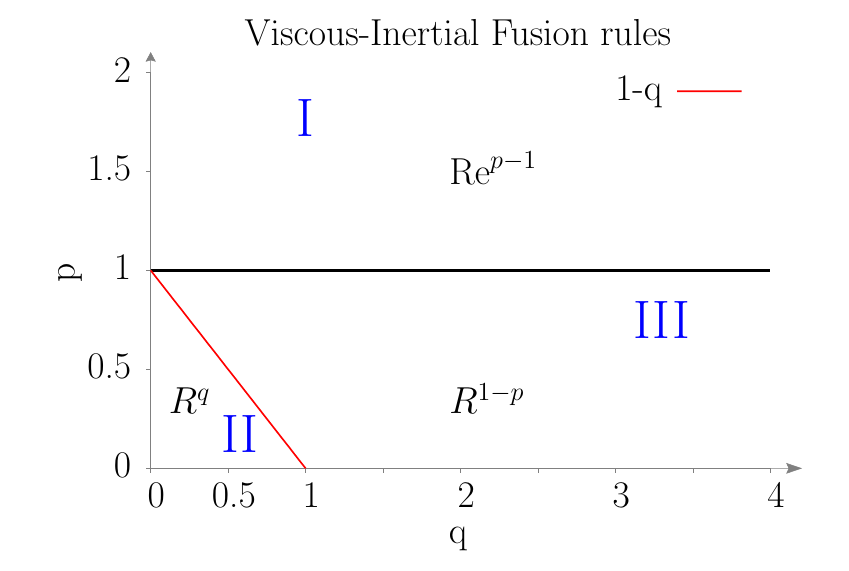}
  \caption{Parametric space of the exponents of Eq.~(\ref{eq:d}) and the
 MF prediction for the viscous-inertial fusion rules. Here I--III correspond to the three regions that emerge from the bifractal Burgers scaling (\ref{eq:burgers_scaling}).}
\label{fig:MF_rls_pred}
\end{figure}

In the following, we consider the viscous-inertial multiscale  correlation function given
\begin{equation}
D_{p,q}(\textrm{Re},R) = \langle (\partial_x v)^p (\delta_R v)^q \rangle.
\label{eq:d}
\end{equation}
We specialize to the Burgers case for which the MF prediction is given in Fig.~\ref{fig:MF_rls_pred} and as follows by inspecting Eq.~\eqref{eq:diss_anomaly_mf}:
\begin{equation}
  D_{p,q}(\textrm{Re},R) \sim
  \begin{cases}
  \, \textrm{Re}^{p-1} & \text{for $p > 1$}\,, \\
  \, R^q & \text{for $p < 1$ \& $p+q \leq 1$}\,, \\
  \, R^{1-p} & \text{for $p < 1$ \& $p+q \geq 1$}\,.
  \end{cases}
  \label{eq:visc_inert_scaling}
\end{equation}
In the following, we will also refer to these relations as the MF prediction for the viscous-inertial fusion rules.

\begin{figure}[t]
\includegraphics[width=0.49 \textwidth]{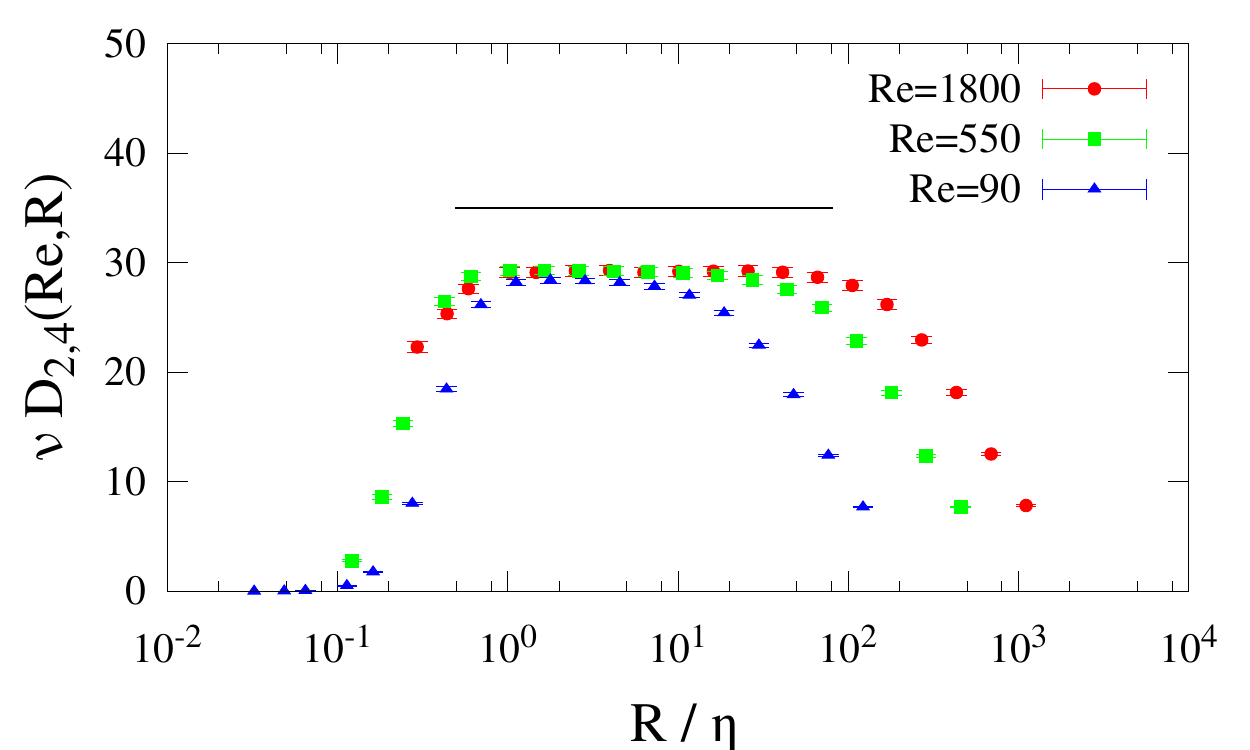}
  \includegraphics[width=0.49 \textwidth]{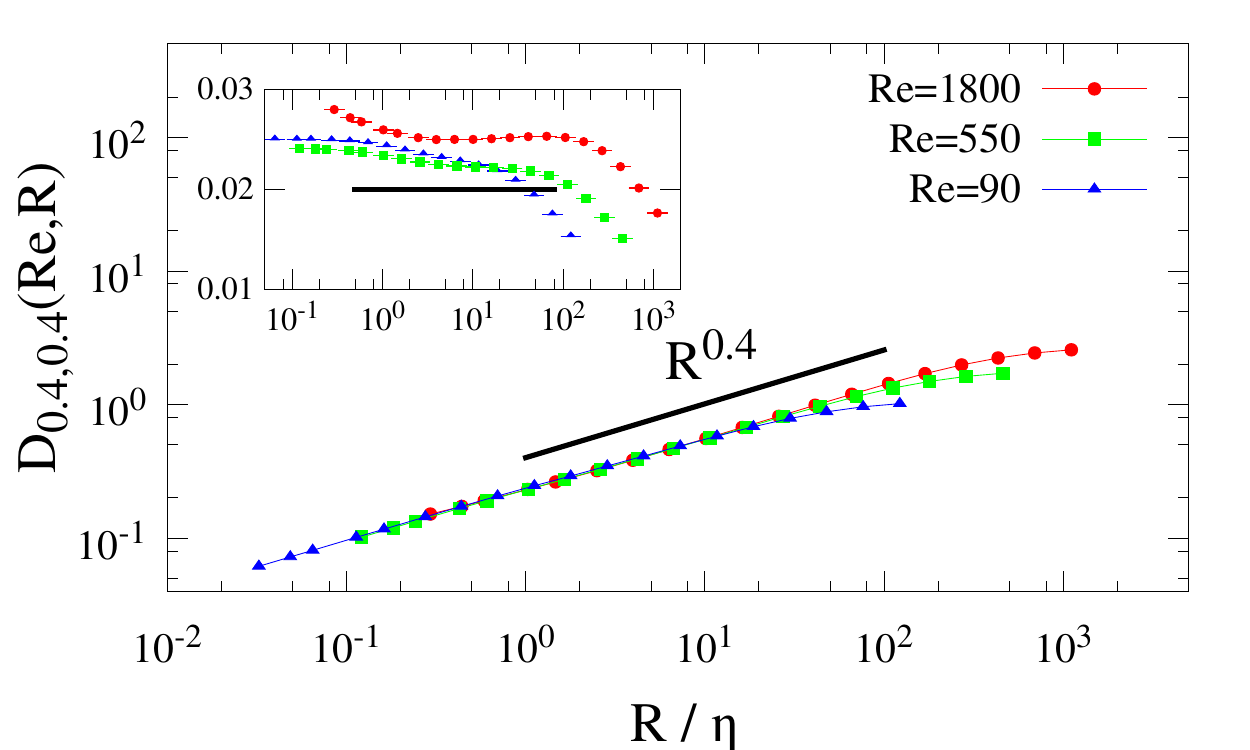}
  \includegraphics[width=0.49 \textwidth]{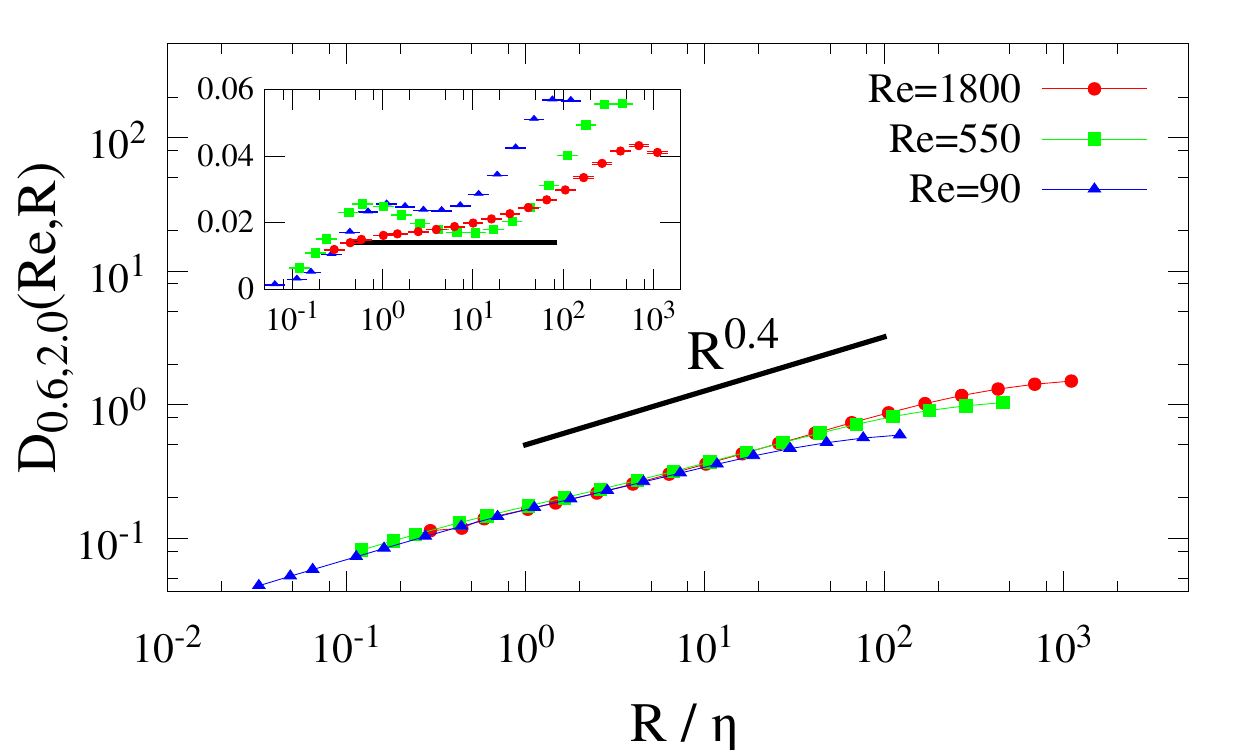}
  \caption{Examination of the viscous-inertial fusion rules prediction \eqref{eq:visc_inert_scaling} via the quantity $D_{p,q}(\textrm{Re},R)$ defined in Eq.~\eqref{eq:d} as a function of $R/\eta$ for different values of $p$ and $q$ and for different Reynolds numbers. The insets depict the results divided by the corresponding viscous-inertial fusion rules prediction of regions I-III of Fig.~\ref{fig:MF_rls_pred}. (a) $p=2$ and $q=4$ (region I). The quantity $\nu \, D_{2,4}(\textrm{Re},R)$ is shown here. It is independent of $R$, and the collapse of the data within errorbars supports the suggestion of a generalized dissipative anomaly discussed in Sec.~\ref{sec:fusion} and Eq.~(\ref{eq:diss_fusion}). (b) $p=q=0.4$ (region II). (c) $p=0.6$ and $q=2$ (region III).}
\label{fig:multifractal}
\end{figure}

Figure~\ref{fig:multifractal}(a) depicts $\nu D_{2,4}(\textrm{Re},R)$ as a function of $R$ for different Reynolds numbers (see also Table \ref{tab:parameters}). As can be seen, $\nu D_{2,4}(\textrm{Re},R)$ is independent of the scale $R$ in the inertial range, which is in accordance with the MF prediction in region I of Fig. \ref{fig:MF_rls_pred}. The collapse of the data comes in agreement with the existence of a generalized dissipative anomaly as discussed in Sec.~\ref{sec:fusion} and Eq.~(\ref{eq:diss_fusion}). The flat region increases as $\nu \rightarrow 0$. Figures~\ref{fig:multifractal}(b) and ~\ref{fig:multifractal}(c) depict $D_{0.4,0.4}(\textrm{Re},R)$ and $D_{0.6,2}(\textrm{Re},R)$ respectively, with combinations of exponents $p$ and $q$ corresponding to regions II and III of Fig.~\ref{fig:MF_rls_pred}, accordingly. The black line with scaling
$R^m$, $m \in \mathbb{R}$, is the viscous-inertial fusion rules prediction following Eq.~(\ref{eq:visc_inert_scaling}) for the chosen exponents $p$ and $q$. The inset depicts the data divided by $R^m$. All three cases show a good agreement of the data with the viscous-inertial fusion rules prediction, which improves as we increase Re.

\subsubsection{Evaluation of the velocity gradient statistics from DNS of Burgers turbulence}
Finally, we want to consider the special case where Eq.~(\ref{eq:d}) reduces to the ordinary moments of the velocity gradient, i.e., $D_{p,0}(\textrm{Re})$. As can be seen from Fig.~\ref{fig:MF_rls_pred}, the MF prediction for Burgers turbulence reduces to the Reynolds number scaling
\begin{equation}
 D_{p,0} \sim \textrm{Re}^{p-1} \qquad \textrm{for $p>1$}\;.
 \label{eq:d_grad}
\end{equation}
Moreover, for the particular case of Burgers turbulence with $\zeta(n)=1$, both
the MF prediction (\ref{eq:diss_anomaly_mf}) and the result from~\cite{yakhot:2001,Yakhot2004,Yakhot2007} in Eq.~(\ref{eq:yakhot_gradient}) yield the relation (\ref{eq:d_grad}), which was already discussed in Sec.~\ref{sec:fusion}.
It is convenient to introduce the quantity
\begin{equation}
  M_{p} = \frac{\left \langle \left(\frac{\partial v}{\partial x} \right)^p \right \rangle}{\left \langle \left(\frac{\partial v}{\partial x} \right)^2 \right \rangle^{p/2}}
  \label{eq:M_n}
\end{equation}
for even $p$. Recent numerical investigations of hydrodynamic turbulence~\cite{Yakhot2017} suggest that the moments (\ref{eq:M_n}) exhibit a transition from Gaussian to anomalous behavior if one increases the Reynolds number. Hence, we expect $M_p$ to behave according to
\begin{equation}
  M_{p} \sim
 \begin{cases}
 \, (p-1)!! & \textrm{ for}\;\textrm{Re} \sim O(1)\;, \\
 \,  \textrm{Re}^{p/2-1} \sim \textrm{Re}_{\lambda}^{p-2}  & \textrm{ for } \textrm{Re} \gg O(1)
 \end{cases}
 \label{eq:M_n_pred}
\end{equation}
for even $p$.
Here we made use of the fact that the Taylor-Reynolds number
$\textrm{Re}_{\lambda}=u_{\rm rms} \lambda/ \nu$ is related to $\textrm{Re}$ according to $\textrm{Re} \sim \textrm{Re}_{\lambda}^2$ in the high-Reynolds-number regime.

\begin{figure}[b]
  \includegraphics[width=0.49 \textwidth]{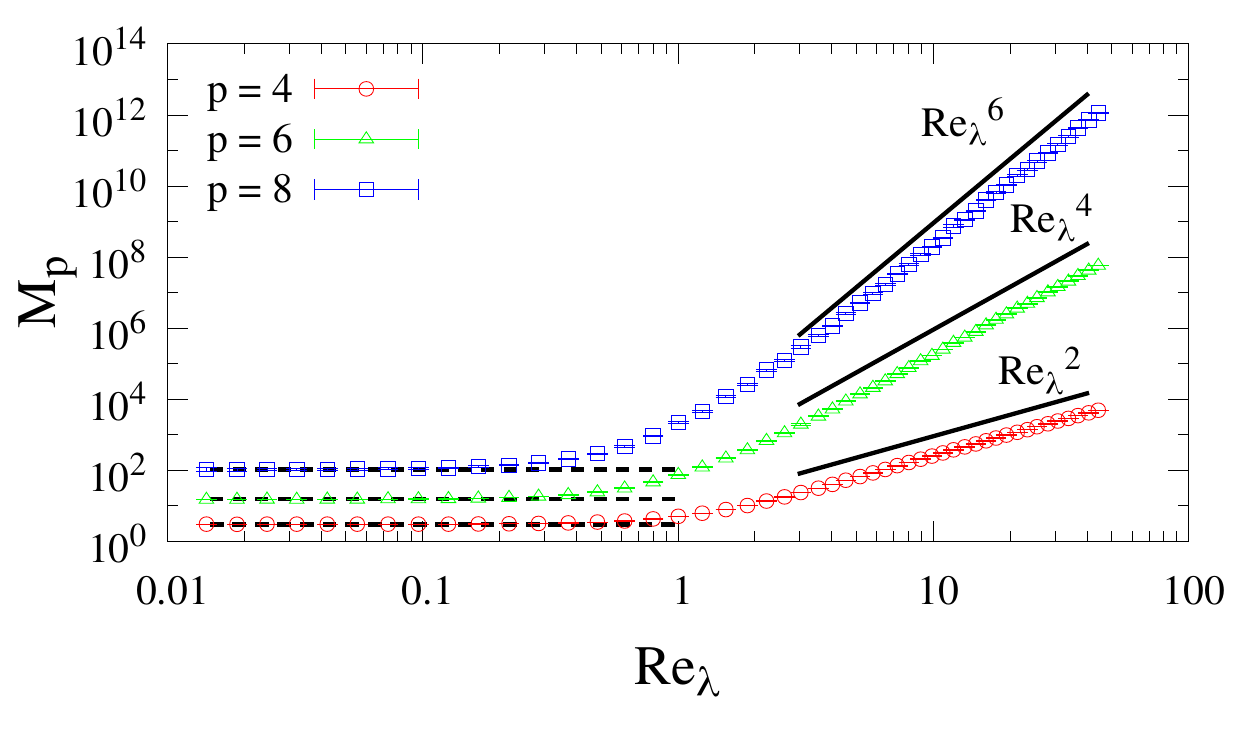}
  \includegraphics[width=0.49 \textwidth]{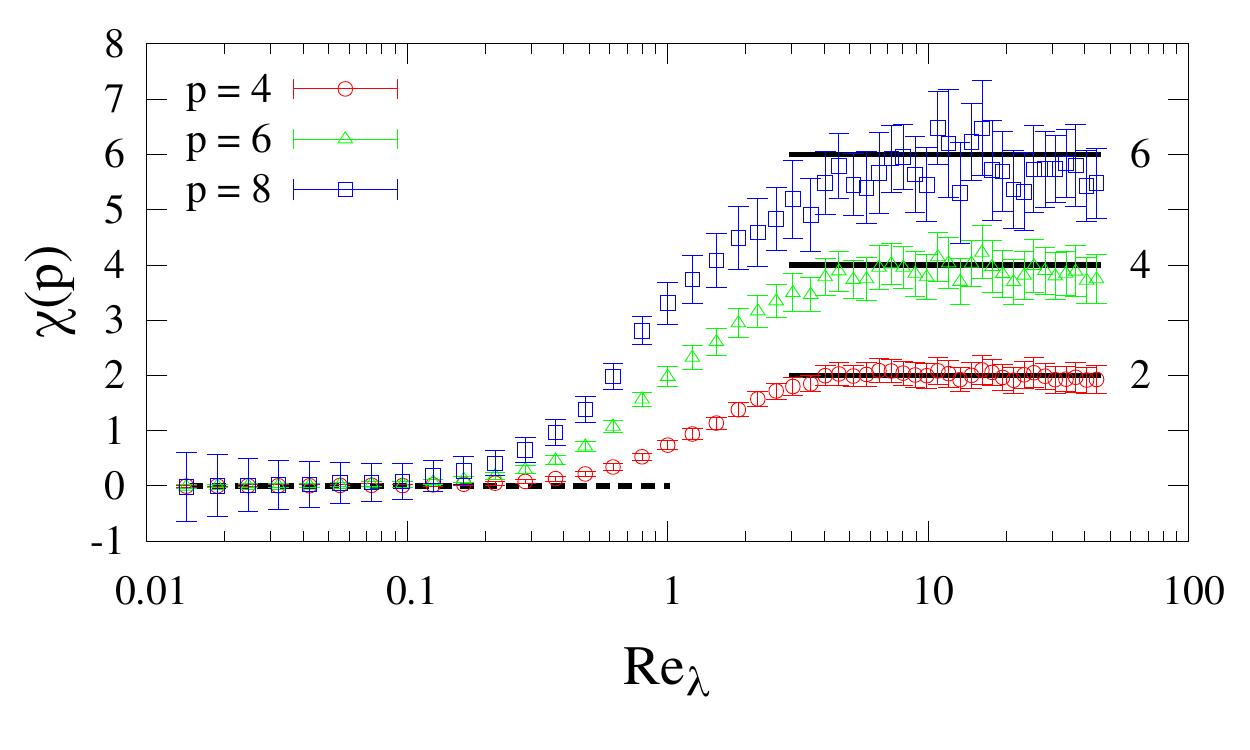}
  \caption{(a) Moments $M_p$ from Eq.~(\ref{eq:M_n}) as a function of the Taylor-Reynolds number in Burgers turbulence. The low-Reynolds-number regime exhibits Gaussian statistics (dashed black lines), whereas the high-Reynolds-number regime agrees well with the multifractal prediction \eqref{eq:diss_anomaly_mf} and the result from \cite{yakhot:2001,Yakhot2004,Yakhot2007} (solid black line) for Burgers turbulence (\ref{eq:M_n_pred}). The lines correspond to flat regions of the logarithmic derivative of the moments $\chi(n)$ [see (b)].
  (b) Logarithmic derivative of the moments (\ref{eq:log_derivatives}). The straight black lines correspond to the theoretical predictions (\ref{eq:M_n_pred}), $\chi(4)=2$, $\chi(6)=4$, and $\chi(8)=6$.}
\label{fig:burgers_transition}
\end{figure}

Figure ~\ref{fig:burgers_transition} is in quantitative agreement with Eq.~(\ref{eq:M_n_pred}). Figure ~\ref{fig:burgers_transition}(a) depicts the moments (\ref{eq:M_n}) as a function of the Taylor-Reynolds number. For small $\textrm{Re}_{\lambda}$, the moments exhibit Gaussian statistics similar to the case of hydrodynamic turbulence~\cite{Yakhot2017}, whereas the anomalous behavior for larger $\textrm{Re}_{\lambda}$
is much more pronounced in comparison to the latter case. Obviously, this result can be attributed to the strong intermittency behavior in Burgers turbulence. Nevertheless, in the high-Reynolds-number regime we can confirm the prediction (\ref{eq:M_n_pred}) to a great extent. The fits (black lines) in Fig.~\ref{fig:burgers_transition}(a) correspond to flat regions in the logarithmic derivative of the moments (\ref{eq:M_n})
\begin{equation}
  \chi(p) = \frac{\textrm{d} \log M_p}{\textrm{d} \log \textrm{Re}_\lambda},
  \label{eq:log_derivatives}
\end{equation}
which is displayed in Fig.~\ref{fig:burgers_transition}(b). The flat regions are indicated as flat lines which correspond to the theoretical predictions (\ref{eq:M_n_pred}), $\chi(4)=2$, $\chi(6)=4$, and $\chi(8)=6$. Hence, we can conclude that the MF prediction also applies to the single-gradient statistics in Burgers turbulence.

\section{Conclusion}
We have presented an overview of prevalent concepts that allow for multiscale descriptions of turbulent flows. A main result of this paper is that the operator-product-expansion--fusion-rules approach~\cite{Eyink1993,Polyakov1995a,Lvov1996,Lvov1996a} that emanated from quantum field theory is a direct consequence of the Markov property of velocity increments in scale devised in~\cite{Friedrich1997}, provided that the structure functions exhibit scaling in the inertial range. This means an amalgamation of two fields that co-existed for nearly 20 years. By contrast, our results might also lead to a novel stochastic interpretation of the operator-product expansion in quantum field theory~\cite{Wilson1969}. Different from other closure methods, e.g.,
the quasinormal approximation~\cite{Millionschikov1941}, renormalization methods~\cite{Kraichnan1957,Wyld1961b}, renormalization-group methods~\cite{Forster1977a}, and the
eddy-damped quasinormal approximation~\cite{Orszag1970,Orszag1974,McCo,Lesieur}, both the Markov approach and the operator-product expansion
are nonperturbative,  i.e.,
are not based on properties of Gaussian-distributed velocity field fluctuations. The latter property makes both approaches suitable candidates for a closure of the multi-increment PDF hierarchy~\cite{Friedrich2017}.

Regarding the breakdown of the fusion rules in the limit of small-scale separations, it is tempting to investigate the influence of non-Markovian cascade processes. Here, a generalization of the KM expansion for the transition PDF
(\ref{eq:km_exp_trans}) to arbitrary stochastic processes as emphasized in \cite{Srinivas1977}, might yield a generalization of the fusion rules to arbitrary cascade processes. A dissipative cutoff of the structure functions~\cite{Meneveau1996} can also be achieved by a dissipative KM expansion, but is beyond the scope of
the present work. In addition, the Markov property could be considered as a first step in an approximation of multi-increment statistics. The natural next step would be an extension incorporating one additional level of memory in scale~\cite{Friedrich2017}, e.g., assuming
\begin{equation}
	p(v_4,r_4|v_3,r_3;v_2,r_2;v_1,r_1) \approx p(v_4,r_4|v_3,r_3;v_2,r_2)\;,
\end{equation}
and thus allowing one to capture correlations between the inertial and viscous-inertial ranges.

Furthermore, we have shown that a specific prediction of the MF model for joint velocity gradient and velocity increment statistics (\ref{eq:diss_fusion}) can be obtained from the basic fluid dynamical equations under the neglect of pressure contributions, i.e., from the Burgers equation. It must be stressed that this result can be derived without any further assumptions apart from the scaling of structure functions in the inertial range. However, at this point, we could not validate the generalization of the MF result to arbitrary powers of the velocity gradient given by Eq.~(\ref{eq:diss_fusion_generalized}). In order to derive such a generalization, one has to operate at the next level of the multi-increment hierarchy (\ref{eq:evo1}). Here a possible closure is the Markov property (\ref{eq:markov}) which, leads to a self-consistent equation for the two-increment PDF~\cite{Friedrich2017}.

The numerical part of this work was devoted to the verification of fusion rules and the prediction of the MF prediction in DNSs of Burgers turbulence. Both fusion rules and MF prediction could be established to a certain extent. The limitation of the fusion rules arises for vanishing scale separations and could be understood from the violation of the Markov property (\ref{eq:markov}). A further examination of this regime is a task left to future research.

\section{Acknowledgement}
J.F. and R.G. acknowledge fruitful discussions with J.~Schumacher and J.~Peinke. L.B. acknowledges funding from the European Research Council under the European Union's Seventh Framework Programme, ERC Grant Agreement No. 339032. G.M. kindly acknowledges funding from the European Union's Horizon 2020 research and innovation program under the Marie Sk\l{}odowska-Curie Grant Agreement No. 642069 (European Joint Doctorate program ``HPC-LEAP'').

\newpage

\appendix

\section{Generalization of fusion rules to $n$-increment statistics}
\label{app:gen_fusion}
We consider the moments of the $n$-increment PDF
\begin{eqnarray}
  \label{eq:n_inc_mom}
  \lefteqn{\left \langle (\delta_{r_n}v)^{p_n} \cdots (\delta_{r_2}v)^{p_2}(\delta_{r_1}v)^{p_1} \right \rangle}\\ \nonumber
  &=&
  \int \textrm{d}v_n \cdots \textrm{d}v_2\; \textrm{d}v_1\;  v_n^{p_n}\cdots v_2^{p_2}\; v_1^{p_1}\; f_n(v_n,r_n;\cdots;v_1,r_1)\;,
\end{eqnarray}
where the $p_i$'s denote arbitrary exponents and we impose the scale ordering
$\eta\le r_n \le r_{n-1}\le \ldots \le r_2 \le r_1 \le L$.
First, we rewrite the $n$-increment PDF according to Bayes' theorem
\begin{eqnarray}\nonumber
  &~&f_n(v_n,r_n;\ldots v_1,r_1)\\ \nonumber
  &=&p(v_n,r_n|v_{n-1},r_{n-1};\ldots; v_1,r_1)\\
  &~&\times f_{n-1}(v_{n-1},r_{n-1};\ldots v_1,r_1).
\end{eqnarray}
The general form of the Markov property in scale implies that
\begin{equation}
  p(v_n,r_n|v_{n-1},r_{n-1};\ldots; v_1,r_1)=
  p(v_n,r_n|v_{n-1},r_{n-1})\;.
\end{equation}
Hence, Eq.~(\ref{eq:n_inc_mom}) simplifies to
\begin{eqnarray}\nonumber
  &~&\int \textrm{d}v_n \ldots \textrm{d}v_2\; \textrm{d}v_1\;  v_n^{p_n}\ldots v_2^{p_2}\; v_1^{p_1}\; p(v_n,r_n|v_{n-1},r_{n-1})\\ &~&\times f_{n-1}(v_{n-1},r_{n-1};\ldots;v_1,r_1)
  \label{eq:n_inc_mom_markov}
\end{eqnarray}
Under the assumption of the scaling of structure functions in combination with the Markov property, we can express the conditional probability $p(v_n,r_n|v_{n-1},r_{n-1})$ in terms of a Dyson series (\ref{eq:dyson})
\begin{widetext}
\begin{eqnarray}\nonumber
  p(v_n,r_n|v_{n-1},r_{n-1})&=& \delta(v_n-v_{n-1}) +\ln \frac{r_{n-1}}{r_n} \sum_{k=1}^\infty \frac{K_k}{k!}
        \frac{\partial^k}{\partial v_n^k} v_n^k \delta(v_n-v_{n-1})\\
  &~&+ \frac{1}{2!} \left(\ln \frac{r_{n-1}}{r_n} \right)^2
  \sum_{k=1}^\infty \frac{K_k}{k!}
        \frac{\partial^k}{\partial v_n^k} v_n^k
        \sum_{l=1}^\infty \frac{K_l}{l!}
              \frac{\partial^{l}}{\partial v_n^{l}} v_n^{l} \delta(v_n-v_{n-1}) + \cdots
\label{eq:dyson_n}
\end{eqnarray}
\end{widetext}
Inserting (\ref{eq:dyson_n}) into (\ref{eq:n_inc_mom_markov})
and performing the partial integrations with respect to $v_n$ similar to Eq.~(\ref{eq:fusion_markov}) yields
\begin{widetext}
  \begin{eqnarray}
    &~&\left[1+ \sum_{k=1}^{p_n} (-1)^{k} K_k{ p_n \choose k}\ln \frac{r_{n-1}}{r_n}+ \frac{1}{2!} \sum_{k=1}^{p_n} (-1)^{k} K_k{ p_n \choose k} \sum_{l=1}^{p_n}
     (-1)^{l} K_l{ p_n \choose l} \left(\ln \frac{r_{n-1}}{r_n} \right)^2 +\ldots \right]\\ \nonumber
    &~&\times\int \textrm{d}v_{n-1} \ldots \textrm{d}v_2\; \textrm{d}v_1\;  v_{n-1}^{p_n+p_{n-1}}\ldots v_2^{p_2}\; v_1^{p_1}\; f_{n-1}(v_{n-1},r_{n-1};\ldots;v_1,r_1)\;.
  \end{eqnarray}
\end{widetext}
Here the term in square brackets can be written as an exponential function according to
\begin{equation}
  \Big [\ldots \Big]=\exp \left[\ln \frac{r_{n-1}}{r_n} \sum_{k=1}^{p_n} (-1)^{k} K_k{ p_n \choose k} \right] \;.
\end{equation}
The sum in the exponential function can be identified as the scaling exponent  $\zeta(p_n)=-\sum_{k=1}^{p_n} (-1)^{k} K_k{ p_n \choose k}$ and we obtain
  \begin{eqnarray}\nonumber
    &~&\left \langle (\delta_{r_n}v)^{p_n} (\delta_{r_{n-1}}v)^{p_{n-1}} \cdots (\delta_{r_2}v)^{p_2}(\delta_{r_1}v)^{p_1} \right \rangle \\ \nonumber
    &=&\exp \left[\ln \frac{r_{n}}{r_{n-1}} \zeta(p_n) \right] \\
    &~& \times \left \langle (\delta_{r_{n-1}}v)^{p_n+p_{n-1}}
      \cdots (\delta_{r_2}v)^{p_2}(\delta_{r_1}v)^{p_1} \right \rangle.
      \label{eq:n_inc_mom_simplified}
  \end{eqnarray}
Furthermore, the scaling of the structure functions implies that $
\left \langle (\delta_{r_n} v)^{p_n} \right \rangle \sim r_n^{\zeta(p_n)}$, which yields
\begin{eqnarray}
  &~&\left \langle (\delta_{r_n}v)^{p_n} (\delta_{r_{n-1}}v)^{p_{n-1}} \cdots (\delta_{r_2}v)^{p_2}(\delta_{r_1}v)^{p_1} \right \rangle\\ \nonumber
  &=&\frac{\left \langle (\delta_{r_n} v)^{p_n} \right \rangle }{\left \langle (\delta_{r_{n-1}} v)^{p_n} \right \rangle }
  \left \langle (\delta_{r_{n-1}}v)^{p_n+p_{n-1}}
   \cdots (\delta_{r_2}v)^{p_2}(\delta_{r_1}v)^{p_1} \right \rangle\;.
\end{eqnarray}
Successive application of this relation yields
\begin{eqnarray}\nonumber
  &~&\left \langle (\delta_{r_n}v)^{p_n} (\delta_{r_{n-1}}v)^{p_{n-1}} \ldots (\delta_{r_2}v)^{p_2}(\delta_{r_1}v)^{p_1} \right \rangle\\ \nonumber
  &=&\frac{\left \langle (\delta_{r_n} v)^{p_n} \right \rangle }{\left \langle (\delta_{r_{n-1}} v)^{p_n} \right \rangle }\times
  \frac{\langle \left (\delta_{r_{n-1}} v)^{p_n+p_{n-1}} \right \rangle }{\left \langle (\delta_{r_{n-2}} v)^{p_n+p_{n-1}} \right \rangle }\times \cdots \\ &~&\times
  \frac{\left \langle (\delta_{r_2} v)^{p_n+\cdots +p_2} \right \rangle }{\left \langle (\delta_{r_{1}} v)^{p_n+\cdots + p_2} \right \rangle }\times
  \left \langle (\delta_{r_1}v)^{p_n + \cdots +p_1} \right \rangle\;,
\end{eqnarray}
or in more compact notation
\begin{eqnarray}
&~&\prod_{i=1}^{n}  \left \langle (\delta_{r_i}v)^{p_i}\right \rangle \\ \nonumber &=&\prod_{i=1}^{n-1} \frac{\left \langle (\delta_{r_{i+1}}v)^{\sum_{k=1}^{i} p_{n+1-k}} \right \rangle}
{\left \langle (\delta_{r_i}v)^{\sum_{k=1}^{i}p_{n+1-k}} \right \rangle}
\left \langle (\delta_{r_1}v)^{\sum_{k=1}^{n}p_{n+1-k}} \right \rangle\;,
\end{eqnarray}
which is the counterpart to Eq.~(\ref{eq:gen_fusion}).

\section{Derivation of multi-increment hierarchy in Burgers turbulence}
\label{app:1}
In order to derive the evolution equation~(\ref{eq:evo1}) we take the temporal derivative of the one-increment PDF
\begin{widetext}
\begin{eqnarray}
 &~&\frac{\partial}{\partial t} f_1(v_1,r_1,t)=
   \frac{\partial}{\partial t}  \langle \delta(v_1-\delta_{r_1} v(x,t)) \rangle =-\frac{\partial}{\partial v_1} \left \langle  \delta(v_1-\delta_{r_1} v(x,t))
      \frac{\partial}{\partial t}  \delta_{r_1} v(x,t) \right \rangle \\ \nonumber
 &=& \frac{\partial}{\partial v_1} \left \langle  \delta(v_1-\delta_{r_1} v(x,t))
\left[v(x,t) \frac{\partial}{\partial x}\delta_{r_1} v(x,t)
+\delta_{r_1} v(x,t)\frac{\partial}{\partial r_1}\delta_{r_1} v(x,t)-\nu\frac{\partial^2}{\partial x^2}\delta_{r_1} v(x,t)
-F(x+r_1,t)+F(x,t) \right] \right \rangle,
\label{eq:evo}
\end{eqnarray}
\end{widetext}
where Eq.~(\ref{eq:burgers_inc}) was used in order to replace the temporal evolution of the velocity increment.
Each term can now be treated separately. Starting with the first advective term, we obtain
\begin{widetext}
\begin{eqnarray}\nonumber
 &-&\frac{\partial}{\partial v_1} \left \langle  \delta(v_1-\delta_{r_1} v(x,t))
v(x,t) \frac{\partial}{\partial x}\delta_{r_1} v(x,t) \right \rangle
= \left \langle v(x,t)\frac{\partial}{\partial x}
\delta(v_1-\delta_{r_1} v(x,t))\right \rangle
= \underbrace{\frac{\partial}{\partial x}\left \langle v(x,t)
\delta(v_1-\delta_{r_1} v(x,t))\right \rangle}_{=0,\; \textrm{homogeneity}}\\
&~&-\left \langle \underbrace{\frac{\partial v(x,t)}{\partial x}
\delta(v_1-\delta_{r_1} v(x,t))}_{=\left[\frac{\partial \delta_{r_1} v(x,t)}{\partial r_1}-
 \frac{\partial \delta_{r_1} v(x,t)}{\partial x}\right]\times \delta}\right \rangle
= \int_{-\infty}^{v_1} \textrm{d}v_1' \frac{\partial}{\partial r_1}
\langle \underbrace{\delta(v_1'-\delta_{r_1} v(x,t)) \rangle}_{=f_1(v_1',r,t)} -\int_{-\infty}^{v_1} \textrm{d}v_1' \underbrace{\frac{\partial}{\partial x}
\langle \delta(v_1'-\delta_{r_1} v(x,t)) \rangle}_{=0,\; \textrm{homogeneity}}\;.
\end{eqnarray}
\end{widetext}
Here we made use of the inverse chain rule in the first and last steps. The second advective term can be treated in the same way according to
\begin{widetext}
\begin{eqnarray}\nonumber
  \lefteqn{-\frac{\partial}{\partial v_1} \left \langle  \delta(v_1-\delta_{r_1} v(x,t))
\delta_{r_1} v(x,t) \frac{\partial}{\partial r_1}\delta_{r_1} v(x,t) \right \rangle}\\ \nonumber
&=& \left \langle \delta_{r_1} v(x,t)\frac{\partial}{\partial r_1}
\delta(v_1-\delta_{r_1} v(x,t))\right \rangle = \frac{\partial}{\partial r_1} \left \langle \delta_{r_1} v(x,t)
\delta(v_1-\delta_{r_1} v(x,t))\right \rangle -\left \langle \frac{\partial \delta_{r_1} v(x,t)}{\partial r_1}
\delta(v_1-\delta_{r_1} v(x,t))\right \rangle \\
&=& v_1 \frac{\partial}{\partial r_1}
\underbrace{\left \langle \delta(v_1-\delta_{r_1} v(x,t))\right \rangle}_{=f_1(v_1,r_1,t)}
+ \int_{-\infty}^{v_1} \textrm{d}v_1' \frac{\partial}{\partial r_1}
\langle \underbrace{\delta(v_1'-\delta_{r_1} v(x,t)) \rangle}_{=f_1(v_1',r_1,t)},
\end{eqnarray}
\end{widetext}
where we made use of the sifting property of the $\delta$ function, i.e.,
$ \delta_{r_1} v(x,t)\delta(v_1-\delta_{r_1} v(x,t))= v_1
\delta(v_1-\delta_{r_1} v(x,t))$. The nonlinear terms can thus be expressed solely in terms of the one-increment PDF or its associated cumulative PDF. which is a particularity of the Burgers equation (for the Navier-Stokes equation we would be facing unclosed terms from the pressure~\cite{Ulinich1969a}).
However, the viscous contributions in Eq.~(\ref{eq:evo}) confront us with unclosed terms and  we have to introduce the two-increment PDF, which results in an infinite hierarchy of PDF equations. This can be seen from the  calculation of the viscous term in Eq.~(\ref{eq:evo}),
\begin{widetext}
\begin{eqnarray}\nonumber
 &-&  \nu \left \langle  \delta(v_1-\delta_{r_1} v(x,t))
 \frac{\partial^2 \delta_{r_1} v(x,t)}{\partial x^2} \right \rangle=-\nu  \left \langle  \delta(v_1-\delta_{r_1} v(x,t))
 \left[\frac{\partial^2 \delta_{r_1} v(x,t)}{\partial r_1^2} -\frac{\partial^2 v(x,t)}{\partial x^2} \right]
 \right \rangle\\ \nonumber
 &=& -\nu \int \textrm {d} r_2
 \left[ \delta(r_2-r_1) - \delta(r_2) \right] \frac{\partial^2}{\partial r_2^2}
 \left \langle \delta_{r_2} v(x,t)  \delta(v_1-\delta_{r_1} v(x,t))   \right \rangle\\
 &=&  -\nu \int \textrm {d} r_2
 \left[ \delta(r_2-r_1) - \delta(r_2) \right] \frac{\partial^2}{\partial r_2^2}
 \int \textrm{d} v_2 v_2
 \underbrace{\left \langle \delta(v_2-\delta_{r_2} v(x,t))
 \delta(v_1-\delta_{r_1} v(x,t))   \right \rangle}_{=f_2(v_2,r_2;v_1,r_1,t)}.
\end{eqnarray}
\end{widetext}
The forcing contributions in Eq.~(\ref{eq:evo}) can be handled by the usual trick of the Langevin
equation. Inserting the above calculations yields the evolution equation for the one-increment PDF
\begin{widetext}
\begin{eqnarray}\nonumber
&~& \frac{\partial}{\partial t} f_1(v_1,r_1,t)+ v_1 \frac{\partial}{\partial r_1}
f_1(v_1,r_1,t)+2\int_{- \infty}^{v_1} \textrm{d} v_1' \frac{\partial}{\partial r_1} f_1(v_1',r_1,t)\\
 &=&
 -\nu \frac{\partial}{\partial v_1} \int \textrm{d} r_2 \left[ \delta(r_2-r_1) -
 \delta(r_2) \right] \frac{\partial^2}{\partial r_2^2}
 \int \textrm{d} v_2 v_2 f_2(v_2,r_2;v_1,r_1,t)
  +[\chi(0)-\chi(r_1)] \frac{\partial^2 }{\partial v_1^2}f_1(v,r_1,t)\;.
  \label{eq:app_evo1}
\end{eqnarray}
\end{widetext}

\section{Reformulation of the viscous term in the multi-increment hierarchy}
\label{app:2}
In this appendix, we show that the unclosed term in the evolution equation of the one-increment PDF (\ref{eq:evo1}) involves the local energy dissipation rate.
To this end, we rewrite the viscous contributions in their original form according to
\begin{eqnarray}\nonumber
 &\nu& \int \textrm{d} r_2 \left[ \delta(r_2-r_1) -
 \delta(r_2) \right] \frac{\partial^2}{\partial r_2^2} \int \textrm{d}v_2 v_2 f_2(v_2,r_2;v_1,r_1)
 \\
&=&  \nu \left \langle \left[\frac{\partial^2 \delta_{r_1} v(x) }{\partial r_1^2}
-\frac{\partial^2 u(x) }{\partial x^2} \right] \delta(v_1-\delta_{r_1} v(x))\right \rangle \;.
 \label{eq:cont}
 \end{eqnarray}
 A further treatment of these terms yields
\begin{widetext}
\begin{eqnarray}\nonumber
 &~&+\nu \frac{\partial}{\partial r_1} \left \langle
 \frac{\partial \delta_{r_1} v(x) }{\partial r_1} \delta(v_1-\delta_{r_1} v(x)) \right \rangle
 -\nu \left \langle
 \frac{\partial \delta_{r_1} v(x) }{\partial r_1}
 \frac{\partial \delta(v_1-\delta_{r_1} v(x))}{\partial r_1} \right \rangle\\ \nonumber
 &~&-\nu \underbrace{\frac{\partial}{\partial x} \left \langle
 \frac{\partial v(x) }{\partial x} \delta(v_1-\delta_{r_1} v(x)) \right \rangle}_{=0, ~\textrm{homogeneity}}
  +\nu \left \langle
 \frac{\partial v(x) }{\partial x} \frac{\partial \delta(v_1-\delta_{r_1} v(x))}{\partial x} \right \rangle\\
 \nonumber
 &=& -\nu  \int_{-\infty}^{v_1} \textrm{d} v_1' \frac{\partial^2}{\partial r_1^2}
 \langle \delta(v_1'-\delta_{r_1} v(x)) \rangle
 +\nu \frac{\partial}{\partial v_1} \left \langle \left(\frac{\partial \delta_{r_1} v(x)}{\partial r_1} \right)^2
 \delta(v_1-\delta_{r_1} v(x)) \right \rangle\\
 &~& -\nu \frac{\partial}{\partial v_1}\left \langle
  \frac{\partial v(x) }{\partial x}
 \underbrace{\left(\frac{\partial \delta_{r_1} v(x) }{\partial x}\right)}_{=\frac{\partial \delta_{r_1} v(x) }
 {\partial r_1}-\frac{\partial v(x) }{\partial x}}\delta(v_1-\delta_{r_1} v(x)) \right \rangle.
\end{eqnarray}
\end{widetext}
Inserting the one-increment PDF $f_1(v_1',r_1)$ into the first term on the right-hand side yields
\begin{widetext}
\begin{eqnarray}\nonumber
 && -\nu  \int_{-\infty}^{v_1} \textrm{d} v_1'
 \frac{\partial^2}{\partial r_1^2} f_1(v_1',r_1)
 +\nu \frac{\partial}{\partial v_1} \left \langle \left(\frac{\partial v(x+r_1)}{\partial r_1} \right)^2
 \delta(v_1-\delta_{r_1} v(x)) \right \rangle\\ \nonumber
 &~&+\nu \frac{\partial}{\partial x}
 \left \langle \underbrace{\left(\frac{\partial v(x)}{\partial x}\right)}_{= \frac{\partial \delta_{r_1} v(x) }
 {\partial r_1}-\frac{\partial \delta_{r_1} v(x)}{\partial x}} \delta(v_1-\delta_{r_1} v(x)) \right\rangle
 +\nu \frac{\partial}{\partial v_1}\left \langle
 \left (\frac{\partial v(x) }{\partial x}\right)^2
  \delta(v_1-\delta_{r_1} v(x)) \right \rangle\\ \nonumber
   &=& -\nu  \int_{-\infty}^{v_1} \textrm{d} v_1'
   \frac{\partial^2}{\partial r_1^2} f_1(v_1',r_1)
 +\nu \frac{\partial}{\partial v_1} \left \langle \left(\frac{\partial v(x+r_1)}{\partial r_1} \right)^2
 \delta(v_1-\delta_{r_1} v(x)) \right \rangle\\
 &~&-\nu \int_{-\infty}^{v_1} \textrm{d}v_1' \frac{\partial^2}{\partial r_1^2}
 f_1(v_1',r)
 +\nu  \int_{-\infty}^{v_1} \textrm{d} v_1'
 \underbrace{\frac{\partial^2}{\partial r_1 \partial x} f_1(v_1',r_1)}_{=0, ~\textrm{homogeneity}}
 +\nu \frac{\partial}{\partial v_1}\left \langle
 \left (\frac{\partial v(x) }{\partial x}\right)^2
  \delta(v_1-\delta_{r_1} v(x)) \right \rangle.
\end{eqnarray}
\end{widetext}
Under the assumption of homogeneity, we obtain
\begin{eqnarray}\nonumber
&~&\left \langle \left(\frac{\partial v(x+r_1)}{\partial r_1} \right)^2
 \delta(v_1-\delta_{r_1} v(x)) \right \rangle \\ \nonumber
 &=& \left \langle \left(\frac{\partial v(x+r_1)}{\partial r_1} \right)^2
 \delta(v_1-v(x+r_1)+v(x)) \right \rangle \\ \nonumber
&=&  \left \langle \left(\frac{\partial v(x)}{\partial x} \right)^2
 \delta(v_1-v(x)+v(x-r_1)) \right \rangle \\
&=&   \left \langle \left(\frac{\partial v(x)}{\partial x} \right)^2
 \delta(v_1+\delta_{-r_1}v(x)) \right \rangle,
\end{eqnarray}
which allows us to introduce the local energy dissipation rate in Eq.~(\ref{eq:evo1}) according to
\begin{eqnarray}\nonumber
 &~&v_1 \frac{\partial}{\partial r_1}
f_1(v_1,r_1)
 =  2\int_{- \infty}^{v_1} \textrm{d} v_1' \frac{\partial}{\partial r_1} f_1(v',r_1)
 \\ \nonumber
   &~&~ -\frac{\partial^2 }{\partial v_1^2}\Bigg[ \left \langle \frac{\varepsilon(x)}{2}
   [\delta(v_1-\delta_{r_1} v(x))+ \delta(v_1+\delta_{-r_1} v(x))] \right \rangle \\
  &~& +[\chi(0)-\chi(r_1)]f_1(v_1,r_1) \Bigg]+2\nu \frac{\partial^2}{\partial r_1^2} f_1(v_1,r_1).
   \label{eq:evo1_eps_neu}
\end{eqnarray}

\bibliographystyle{apsrev4-1}
\bibliography{paper}

\end{document}